\documentclass[11pt,letterpaper,fleqn]{article}
\usepackage{color,graphicx,natbib,lscape,here,geometry,setspace,multirow,enumitem}
\usepackage[dvipsnames]{xcolor}
\pdfoutput=1
\usepackage{graphicx}
\onehalfspace

\usepackage{authblk}
\usepackage{silence}
\usepackage{rotating}
\usepackage{multirow}
\usepackage{multicol}

\usepackage{booktabs}
\usepackage{color}
\usepackage{setspace}
\usepackage{algorithm2e}
\usepackage{enumitem}
\usepackage{tikz}

\usepackage{newtxtext,newtxmath}
\newcommand*{\scfont}{\fontfamily{ptm}\selectfont}

\definecolor{nblue}{HTML}{000660}
\usepackage[colorlinks=true,urlcolor=nblue,linkcolor=nblue,citecolor=nblue]{hyperref}
\usepackage{bigstrut}

\usepackage{tabularx}
\usepackage{threeparttable}
\usepackage{dcolumn}

\usepackage{etoolbox} 
\makeatletter
\patchcmd{\BR@backref}{\newblock}{\newblock[}{}{}
\patchcmd{\BR@backref}{\par}{]\par}{}{}
\makeatother

\usepackage{afterpage}

\usepackage{longtable}
\usepackage{array}
\newcolumntype{C}[1]{>{\centering\arraybackslash}p{#1}}

\usepackage{comment}
\usepackage{mathtools}

\usepackage[hang,flushmargin]{footmisc} 
\usepackage[british]{babel}

\usepackage[nolists,tablesfirst]{endfloat}                  
 \usepackage{float}
 \restylefloat{table}

\usepackage[title,titletoc]{appendix}
\makeatletter

\renewenvironment{appendices}{%
    \begin{oldappendices}%
    \renewcommand{\thefigure}{\ifnum \c@section>\z@ \thesection.\fi\@arabic\c@figure}%
    \@addtoreset{figure}{section}%
    \renewcommand{\thetable}{\ifnum \c@section>\z@ \thesection.\fi\@arabic\c@table}%
    \@addtoreset{table}{section}}{%
    \end{oldappendices}%
}\makeatother

\usepackage{titlesec} 
\titleformat{\section}[block]{\large}{\thesection. }{0em}{\MakeUppercase} 
\titleformat{\subsection}[block]{\large}{\thesubsection. }{0em}{\itshape} 
\titleformat{\subsubsection}[block]{\large}{}{0em}{\itshape} 

\let\natbibcitet\citet
\renewcommand\citet{\bibpunct{(}{)}{,}{a}{,}{,}\natbibcitet}
\let\natbibcitep\citep
\renewcommand\citep{\bibpunct{(}{)}{;}{a}{,}{;}\natbibcitep}
\newcommand{\bi}{\begin{itemize}}
\newcommand{\ei}{\end{itemize}}
\newcommand{\be}{\begin{equation}}
\newcommand{\ee}{\end{equation}}
\defcitealias{ieo14}{IEO, 2014}

\long\def\symbolfootnote[#1]#2{\begingroup%
\def\thefootnote{\fnsymbol{footnote}}\footnote[#1]{#2}\endgroup}

\makeatletter
\def\ubar#1{\underline{\sbox\tw@{$#1$}\dp\tw@\z@\box\tw@}}
\def\obar#1{\overline{\sbox\tw@{$#1$}\dp\tw@\z@\box\tw@}}
\makeatother

\usepackage{bm}
\usepackage{caption}
\usepackage{subcaption}

\makeatletter\let\p@subfigure\thefigure\makeatother

\floatstyle{plaintop}
\restylefloat{table}

\usepackage{fancyhdr} 
\usepackage{cleveref}
\crefname{chapter}{Chapter}{Chapters}
\crefname{section}{Section}{Sections}
\crefname{subsection}{Section}{Sections}
\crefname{subsubsection}{Section}{Sections}
\crefname{figure}{Figure}{Figures}
\crefname{table}{Table}{Tables}
\crefname{equation}{Equation}{Equations}
\crefname{appendix}{Appendix}{Appendices}
\crefname{appendices}{Appendix}{Appendices}

\crefname{appsec}{Appendix}{Appendices}

\usepackage{catoptions}
\makeatletter

\def\Autoref#1{%
  \begingroup
  \edef\reserved@a{\cpttrimspaces{#1}}%
  \ifcsndefTF{r@#1}{%
    \xaftercsname{\expandafter\testreftype\@fourthoffive}
      {r@\reserved@a}.\\{#1}%
  }{%
    \ref{#1}%
  }%
  \endgroup
}
\def\testreftype#1.#2\\#3{%
  \ifcsndefTF{#1autorefname}{%
    \def\reserved@a##1##2\@nil{%
      \uppercase{\def\ref@name{##1}}%
      \csn@edef{#1autorefname}{\ref@name##2}%
      \autoref{#3}%
    }%
    \reserved@a#1\@nil
  }{%
    \autoref{#3}%
  }%
}
\makeatother

\newcolumntype{d}[1]{D{.}{.}{#1}}

\title{\LARGE{Predicting credit default probabilities using machine learning techniques in the face of unequal class distributions}}
\author{\large{\uppercase{Anna Stelzer}}\thanks{\textit{Address}: Institute for Economic Geography and GIScience, Department of Socioeconomics, Welthandelsplatz 1, A-1020 Vienna, Austria. \textit{Email}: \href{mailto:anna.stelzer@wu.ac.at}{anna.stelzer@wu.ac.at}. The author gratefully acknowledges Florian Huber for his helpful comments and support.
\textit{Date}: \today.}\\\vspace*{-0.5em}\normalsize{\textit{Vienna University of Economics and Business}}}
\date{}


\def\equationautorefname~#1\null{%
  Eq.~(#1)\null
}
\def\equationautorefname~#1\null{
Eq.~(#1)\null
}

\usepackage[utf8, latin1]{inputenc}                 

\begin{document}
\maketitle\thispagestyle{empty}\normalsize\vspace*{-2em}\small

\begin{center}
\begin{minipage}{0.8\textwidth}
\noindent\small This study conducts a benchmarking study, comparing 23 different statistical and machine learning methods in a credit scoring application. In order to do so, the models' performance is evaluated over four different data sets in combination with five data sampling strategies to tackle existing class imbalances in the data. Six different performance measures are used to cover different aspects of predictive performance. The results indicate a strong superiority of ensemble methods and show that simple sampling strategies deliver better results than more sophisticated ones. \\\\ 
\MakeUppercase{\textit{keywords}}: Forecasting, credit scoring, imbalanced data sets, classification, benchmarking\\
\end{minipage}
\end{center}
\bigskip\normalsize

\renewcommand{\thepage}{\arabic{page}}
\setcounter{page}{1}

\section{Introduction}
\label{sec:intro}

Private household debt is a crucial figure in the analysis of an economy's development and its financial system's stability. 
The amount of household debt, loans and debt securities as a percentage of GDP is still rising in many advanced economies after a small decline during and in the aftermath of the Global Financial Crisis. 
In the United States, this ratio stood at around 79\% in 2016 after reaching its peak in 2007 when household debt, loans and debt securities amounted for almost 98\% of GDP. 
A similar development can be observed in the United Kingdom, Spain and Portugal. 
In countries such as France or Australia, the ratio has been increasing steadily since 2000, with no decrease after the Global Financial Crisis. 
Two exceptions are Germany and Japan, where the figure has decreased since the beginning of the century.

Another important trend is the constant rise in private debt in emerging countries. In China, the amount of household debt, loans and debt securities as a percentage of GDP almost quadrupled between 2006 and 2016 from 11\% to 44\%.
In India, the number increased from less than 3\% in 2000 to almost 10\% in 2016 \citep{IMF2018}. 
Even though these numbers are still low compared to those economically more advanced countries mentioned above, they show that private debt is increasing in significance in some of the largest economies in the world as well as in those that still have a lot of untapped economic potential.

These trends, together with the fact that a large-scale default in private debt has proven to be able to destabilize financial markets considerably during the Global Financial Crisis, emphasize the importance of a reliable and precise evaluation of consumer credit default risks, which is not only crucial for the economic success of individual creditors, but also for global financial stability. 

One way of assessing credit default risk is credit scoring, which is the evaluation of risk associated with giving a loan to an individual or an organization \citep{Crook2007}. 
In order to do so, empirical models estimate the probability that a debtor will behave undesirably in the future, for example if he or she repays installments late or defaults on the loan completely \citep{Lessmann2015}. 
In this sense, all potential borrowers can be categorized into being either ``good'' or ``bad'' customers, where being a good customer implies full repayment until the end of the loan term and being a bad customer implies at least a partly default on the loan. 
Scorecards, tools developed in the credit industry evaluating such risks, provide a mapping that connect the data collected on all existing customers to the binary space of two classes, good and bad. 
The mapping usually uses the information of a loan applicant or borrower to calculate a numeric score.
A low score indicates that the customer is more likely to belong to a certain class, whereas a high score implies that he or she belongs to the other class. 
In a further step, the score is compared to a threshold $\tau$. 
Customers with a score above that threshold are assigned one class, customers with a score below $\tau$ are assigned the other class \citep{Hand2005}.

The use of machine learning methods in the realm of credit scoring has already been widely discussed and a variety of classification techniques has been proposed and assessed in that context \citep[see, for example][]{Baesens2003, Lessmann2015, Nanni2009, Yeh2009}. 
Even early studies on the topic indicated that methods based on machine learning such as a multilayer perceptron perform better in the context of credit scoring and bankruptcy prediction than traditional statistical methods \citep[for an overview, see ][]{Nanni2009, Lessmann2015}.

This paper aims to enrich the existing literature by combining several of foci of the credit scoring literature and evaluating the merits of each method. 
The predictive performance of a variety of classifiers on four different data sets is compared with each other by employing several performance measures frequently used in the credit scoring literature.
Furthermore, different sampling techniques are used and compared with respect to class imbalances in the data, namely, down-sampling, up-sampling, synthetic minority over-sampling technique (SMOTE), BorderlineSMOTE (BSMOTE) and Randomly Over Sampling Examples (ROSE). 
This approach allows to compare models both in relative and absolute terms with respect to class imbalances, that is, whether a technique performs better than others when imbalances exist or whether this relative advantage disappears once the unequal class distributions are handled with suitable data preprocessing methods. 
In addition, it points into the right direction when one is concerned with credit risk predictions in practice.

The remainder of the paper is structured as follows. 
Section \ref{sec:literature} provides a short review of the existing literature. 
Section \ref{sec:algorithms} introduces all classifiers used in the study. 
Section \ref{sec:design} gives an overview of the experimental design, namely, the data sets used, all data preprocessing steps, sampling methods and performance measures. 
The classifiers' performance will be presented and discussed in Section \ref{sec:results}, while the last section summarizes and concludes the paper.

\section{Literature Review}
\label{sec:literature}
While other studies only evaluate few models, \citet{Baesens2003} is one of the first papers comparing classification methods on a larger scale in the credit scoring context. 
17 different individual classifiers are applied on eight different credit scoring data sets in order to give a reliable impression of a classifier's performance, which may vary from data set to data set. 
The authors find that linear classifiers perform only slightly worse than non-linear ones, which points to the fact that only weak non-linearity exists in the data. 
Furthermore, they conclude that most methods performed similarly well and only a few of their tested models are truly inferior to other alternatives.

\citet{Yeh2009} apply methods such as k-nearest neighbors (KNN), logistic regression, discriminant analysis, Naive Bayes, artificial neural networks (ANN) and classification trees on a data set of customers' credit default in Taiwan. 
The goal of the study is to obtain reliable estimates of default probability, and the authors find that ANN is the only technique able to achieve this objective.
While the first two cited studies only consider individual classifiers, \citet{Nanni2009} include ensemble classifiers in their evaluation. 
Ensembles are supposed to improve predictive performance by combining the prediction of multiple individual classifiers, a hypothesis confirmed by the results of \citet{Nanni2009}.

While the choice of classifiers is an important one, credit scoring poses some other task-specific challenges.
\citet{BROWN2012} focus on the issue of imbalanced data sets. 
Loans tend to be granted more readily to customers who are expected to have a low probability of default and considered able to repay the loan in full. 
This circumstance is usually reflected in the data and, consequently, the instances in the available data are not chosen at random. 
Most data sets containing information about borrowers have many more good customers than bad ones.
\citet{BROWN2012} apply ten forecasting techniques to five different credit scoring data sets with varying class frequencies and evaluate their performance with respect to class imbalances.
They find that ensemble methods such as stochastic gradient boosting and random forests perform relatively well in the face of severe class imbalances. 
Furthermore, according to \citet{BROWN2012}, commonly used techniques like linear discriminant analysis (LDA) and logistic regression can compete with more complex methods, but the use of a linear kernel least squares support vector machine is not advisable when very large class imbalances exist in the data.

\citet{Lessmann2015} extend the study by \citet{Baesens2003} by applying individual classifiers as well as homogeneous ensemble methods, which combine predictions of the same kind of individual classifiers, and heterogeneous ensembles, which incorporate predictions of different classifiers, on eight different credit scoring data sets. 
While most benchmarking studies use only one performance measure to compare different classification methods with each other, \citet{Lessmann2015} use a multiplicity of measures that also cover three different aspects of a scorecard.

First, measures such as the AUC, short for area under curve, namely the Receiver Operating Characteristics or ROC curve, or the H-measure assess the discriminatory ability of a classifier. 
Second, metrics like accuracy (i.e., the percentage of correctly classified cases, short PCC) of a classifier is concerned with the correctness of a classifier's categorial prediction. 
Third, measures like the Brier Score evaluate the accuracy of probabilistic predictions. 
The authors conclude that several classifiers predict credit default risk significantly more accurately than the industry standard logistic regression. 
Moreover, they find that some advanced methods perform extremely well, but that many novel scoring techniques cannot improve upon their predecessors, indicating that not all efforts in the development of novel scoring techniques may be worthwhile. 
Additionally, while some of the best performing methods in their study belong to the family of heterogeneous ensembles, the authors recommend using random forest as a benchmark for new model testing, as it is a very robust, well-performing classifier, which is also available in standard software.

\section{Classifiers for credit default predictions}
\label{sec:algorithms}
This section presents the general procedure in estimating a customer's scorecard using the probability of non-default as well as the classification models in this study.

Using classification models, credit scorecards are built by estimating the probability of non-default or default. 
Given a $n$-dimensional vector of a customer $i$'s characteristics $\textbf{x}_i = (x_1, \dots, x_n)' \in \mathbb{R}^n, \quad i=1, \dots, N$, one can estimate the probability of a binary response variable $y_i \in \{0;1\}$ either taking one value or the other, where $y_i=0$ indicates a good customer or no default and $y_i=1$ indicates the opposite, i.e., a defaulting customer.

Now, define $p(y_i=0|\textbf{x}_i)=p(+|\textbf{x}_i)$ as the posterior probability that no default will occur for costumer $i$ and correspondingly, $p(y_i=1|\textbf{x}_i)=p(-|\textbf{x}_i)$.
Once $p(+|\textbf{x}_i)$ has been estimated using one of the methods described below, the estimated probability is compared to a threshold $\tau$ in order to decide whether to label the respective customer or loan as good or bad. 
If $p(+|\textbf{x}_i) \geq \tau$, the customer or loan is considered ``good'', if $p(+|\textbf{x}_i) < \tau$, the label ``bad'' is assigned. 
Following \citet{Lessmann2015}, $\tau$ is calculated for every data set so that the share of instances classified as ``good'' in the test set equals the fraction of ``good'' instances in the training set.

The techniques employed in this study to estimate $p(+|\textbf{x}_i)$ include ten individual models comprising commonly used statistical and machine learning methods. 
Statistical models include logistic regression and discriminant analysis, while machine learning methods include different variants of support vector machines (SVM), decision trees, artificial neural networks (ANN), a k-nearest neighbor classifier (KNN) and Naive Bayes (NB). 
Furthermore, ten homogeneous ensemble models are included, as well as three heterogeneous ensembles, which all combine predictions from the aforementioned individual classifiers. An overview over all 23 models is presented in Table \ref{tab:tab_methods}.

All individual models and homogeneous ensembles were implemented using the \texttt{caret} package in R by \citet{Kuhn2018}, the remaining heterogeneous models are based on customized computer code, using base models implemented in \texttt{caret}. 
The techniques include linear and non-linear models (e.g., linear SVM and SVM with radial or polynomial kernel function), as well as parametric and non-parametric methods (e.g., logistic regression and classification trees), covering several different approaches in the credit scoring literature.
Given the number of models in this study and the fact that all of the employed techniques are well documented elsewhere in the literature, they will not be described in detail here. 
However, as most of the models in use belong to a family or type of techniques, a rough overview over the latter is given in the following.

\subsection{Individual models}

\subsubsection{Logistic Regression}
The logistic regression (LogReg) specifies that an appropriate function of $p(y_i=1|\textbf{x}_i)$ is a linear function of the observed values of the available predictors \citep{Yeh2009}. 
Thus, $p(y_i=1|\textbf{x}_i)$ is specified as
\begin{equation}
    p(y_i=1|\textbf{x}_i) = \frac{1}{1 + exp(-(\alpha + \boldsymbol{\beta}'\textbf{x}_i))},
\end{equation}
where the scalar $\alpha$ is the intercept and $\boldsymbol{\beta}$ is the $n$-dimensional parameter vector. 
Both $\alpha$ and $\boldsymbol{\beta}$ are usually estimated using maximum likelihood methods. 
Logistic regression therefore produces a simple probabilistic formula for classification tasks but has the downside of only capturing linear relationships adequately and not properly considering interaction effects of predictors.

\subsubsection{Discriminant Analysis}
Discriminant Analysis (DA), another standard tool for classification, maximizes the distance between different groups and minimizes the distance within each group. 
In order to do so, DA assigns an instance $\textbf{x}_i$ to a class $y_i \in \{0;1\}$ which has the largest posterior probability $p(y_i|\textbf{x}_i)$. According to Bayes' Rule, the posterior probability is defined as
\begin{equation}
\label{eq:Bayes}
    p(y_i|\textbf{x}_i) = \frac{p(\textbf{x}_i|y_i) p(y_i)}{p(\textbf{x}_i)}.
\end{equation}
DA assumes $p(\textbf{x}_i|y_i)$ to be a multivariate Gaussian with mean vectors $\mu_0$, $\mu_1$ and covariance matrices $\Sigma_0$, $\Sigma_1$.\\
The classification rule then states that $\textbf{x}_i$ is assigned to class 1 if
\begin{equation}
\begin{aligned}
    (\textbf{x}_i - \mu_1)' \Sigma_1^{-1} (\textbf{x}_i - \mu_1) - (\textbf{x}_i - \mu_0)' \Sigma_0^{-1} (\textbf{x}_i - \mu_0) \\
    < 2(\log(p(y_i=1)) - \log(p(y_i=0)))  + \log|\Sigma_0| - \log|\Sigma_1|.
\end{aligned}
\end{equation}
When the two classes are assumed to have different covariance matrices $\Sigma_0$ and $\Sigma_1$, the decision boundary is quadratic in $\textbf{x}_i$ and the classifier is called quadratic discriminant analysis (QDA). 
When the classes have the same covariance matrices, i.e., $\Sigma_0=\Sigma_1$, the decision rule simplifies and becomes linear in $\textbf{x}_i$.
Hence, the classifier is called linear discriminant analysis \citep[LDA,][]{Baesens2003}.
However, the assumption of a Gaussian distribution is rarely justified in classification problems, and while using QDA allows for more generalization, globally quadratic boundaries may also not be suitable for a classification problem at hand. 
Moreover, LDA struggles in the face of large data samples or with many predictors, as the linear boundaries lead to underfitting. 
Therefore, \citet{hastie1994} propose the use of regression procedures to estimate non-linear classification boundaries. 
By using regression methods, the predictors are either expanded or selected in a first step. 
After this basis transformation, a (penalized) LDA is conducted in the new space. 
When choosing a simple linear regression as regression method, as in this study, this procedure results in a more flexible form of LDA, called flexible discriminant analysis (FDA).

\subsubsection{Support Vector Machine}
Support vector machine (SVM) classifiers, first proposed by \citet{Cortes1995}, use hyperplanes as boundaries to divide data into groups of similar class values. 
SVM to implement non-linear class boundaries by mapping input vectors non-linearly into a high-dimensional feature space \citep{kumar2007bankruptcy}. 
Then, the hyperplanes, which are the optimal separating boundaries for the classes, are constructed. 
The maximum margin hyperplane is the one that creates the greatest separation between the classes. 
All training cases closest to it are called support vectors. 
If the data are linearly separable, we speak of a linear SVM (SVM-L). 
However, if the data do not fulfill this requirement of linearity, SVM can use kernels to map a classification problem into a higher dimensional space. 
By doing so, additional dimensions are added to the data so that it becomes separable. 
In this study, a SVM with radial basis kernel function (SVM-R) is used in addition to SVM-L.

\subsubsection{Artificial Neural Networks}
The artificial neural network (ANN) is inspired by biological neural networks of the human nervous system.
Instead of neurons, ANN uses a network of artificial neurons or nodes in order to solve learning problems. 
The nodes of neural networks have one-way connections to other nodes, effectively connecting input and output variables \citep{ripley2007}. 
The nodes can be parallel and interconnected and are arranged in different layers \citep{kumar2007bankruptcy}, which results in great flexibility in the modelling processes.
Thus, ANN can easily handle non-linearity and interaction effects. 
However, one of the disadvantages of these complex structures is that ANN is a black-box model and no simple probabilistic formula for classification can be derived.

\subsubsection{Classification and Regression Tree}
Classification and Regression Trees (CART) utilize a decision-tree-like structure to partition the data space so that cases of the same class will be grouped together. 
Starting at a root node, trees grow by splitting the data upon one variable at each node, resulting in a tree where each branch represents an outcome of those splits.
Each leaf node, that is, the last nodes of a tree, represents a class that is assigned. 
CARTs conduct the data splitting process using one predictor at a time and the splitting decision is based on minimizing heterogeneity in the sample, so that after each split, the resulting subset is more similar than the one before. 
There are different types of classification and decision trees, for example, the C4.5 trees or C5.0 trees by \citet{quinlan1992learning, quinlan2014}. 
The CART technique was developed by \citet{cart84} and is the base model for the random forest classifier. 
One drawback of decision trees in general is that their splitting rules and decisions rely heavily on the data structure, which makes them very sensitive to even small changes in the data.

\subsubsection{k-Nearest Neighbor}
K-Nearest neighbor classifiers (KNN) label instances by assigning them with the class of similarly classified cases. 
The $k$ closest observations to $\textbf{x}_i$, as measured in Euclidean distance, are found in the input space. 
The class of the majority of the $k$-nearest neighbors then determines the class of $\textbf{x}_i$ \citep{Hasti2003}. 
The probability of default is calculated by dividing the number of default cases by the number of the overall nearest neighbors for a chosen $k$. 
The parameter $k$ can be chosen in the tuning process to ensure optimal performance of the model.

\subsubsection{Naive Bayes Classifier}
The Naive Bayes classifier (NB) starts by learning the class-conditional probabilities $p(x_j|y_i)$ for $j=1, \dots, n$, and then proceeds by classifying a new instance by using Bayes' rule, equation \ref{eq:Bayes}, to estimate $p(y_i=0|\textbf{x}_i)$. 
In order to do so, NB assumes class conditional independence, i.e., the effect of an attribute value on a given class is independent of the values of other attributes \citep{Yeh2009}.
This simplifies $p(\textbf{x}_j|y_i)$ such that $p(\textbf{x}_i|y_i) = \prod^n_{j=1} p(x_j|y)$,
where $p(x_j|y)$ are calculated using frequency counts for discrete variables and a normal or kernel based method for continuous variables \citep{Baesens2003}. 
While the assumption of class conditional independence simplifies computation significantly, it also poses a major weakness of NB, as it implies that predictive performance strongly depends on this assumption to hold.

\subsection{Homogeneous ensembles}
As mentioned before, homogeneous ensemble methods combine predictions of multiple base models of the same type.

\subsubsection{Bagging}
Bagging, short for bootstrap aggregating, is a general strategy to boost a model's performance and was proposed by \citet{Breiman1996Bag}. 
It creates a number of new training sets by bootstrap sampling the original training data. 
After the base models are trained on the bootstrap samples, their predictions are then averaged (for continuous target variables) or the class of a test case is assigned according to the bootstrap samples' majority vote (for discrete target variables). 
This approach reduces the predictions' variance \citep{Hasti2003}. 
This simple ensemble method can significantly improve predictive performance, especially for unstable base learners like classification trees. 
As the latter change dramatically with changes in the data, diversity is ensured in the bagging ensemble, which results in predictions that are more robust. 
In this study, CART (Bag-CT) and FDA (Bag-FDA) serve as base models for bagging ensembles.

\subsubsection{Boosting}
Another family of simple ensemble methods, which boost the performance of models with weak predictive performance, is boosting, as proposed by \citet{Schapireetal2012}. 
Similarly to bagging, boosting makes predictions by training models on re-sampled data and the final prediction is determined by a vote of the included base models. 
But while bagging chooses the samples according to bootstrap sampling, boosting chooses the samples in such a way that the base learners trained on those data samples are all weak, but complementary classifiers. 
By doing so, new information is introduced whenever a base model is called and trained on another part of the training data. 
An additional difference to bagging is that the base models do not have an equal vote when choosing the final prediction. Instead, each base model's vote is assigned a weight that is based on its past predictive performance.

Just like bagging, a variety of different boosting algorithms exists, and boosting is not restricted to a specific kind of base model, even though the most commonly used ones are decision trees. This study presents results for boosted classification trees (Boost-CT), boosted logistic regression (Boost-Logit) as well as the following boosting algorithms.

One specific boosting algorithm is AdaBoost by \citet{freund1997}. Given a specific classification problem with a training set and a type of base model, AdaBoost iteratively trains the base models with another training set in each round. 
The training sets are chosen by maintaining a distribution over the training instances, which assigns a weight to each observation. 
This weight measures the significance of predicting the right class of the given training case in the current round. 
While the weights are set equally in the beginning, the weights of incorrectly classified training cases increase with each round. 
This implies that cases that are more difficult to classify receive higher weights effectively focusing more on these cases \citep{Schapireetal2012}.

Gradient boosting is another boosting method that is based on the fact that boosting can be understood as an approach which minimizes an appropriate loss function. 
Gradient boosting \citep{friedman2001} proposes a two-step procedure. 
Starting with a weak base model, an improved model is constructed by fitting a new classification function using least squares with the residuals of the current base model. 
Once this new classification function is formed, it is combined additively with the old base model, effectively resulting in a new, improved model. These steps are repeated for a given number of iterations.

Stochastic gradient boosting \citep[SGB,][]{FRIEDMAN2002} is yet another gradient descent algorithm modified on the grounds of the bagging procedure of \citet{Breiman1996Bag}. 
This modification introduces randomness into the gradient boosting procedure by drawing a random data sample from the training data (without replacement). 
Instead of the full data set, only this subset is used to train base models in each round and update the current iteration. 
\citet{FRIEDMAN2002} finds that this introduction of randomization improves the accuracy of gradient boosting substantially, however, the reason for this is unclear.

\subsubsection{Random Forests}
Random Forests (RF), developed by \citet{Breiman2001}, combine tree models and improve predictions by incorporating the principles of both bagging and boosting. 
Decision trees are trained on bootstrap samples of the training data. 
Unlike in ordinary decision trees, a tree cannot, however, choose upon all variables at a given node. Instead, the tree is generated using random feature selection, which introduces additional diversity. 
After a given number of trees has been grown, the final prediction is generated by collecting all the trees' votes. 
While this procedure makes RF a black box model, it has proven to be a successful and robust ensemble method. 
\citet{Breiman2001} finds that RF are competitive with boosting or adaptive bagging without having to change the training set progressively. 
Furthermore, random input cases and random variable selection produce good predictions, especially in classification problems. 
As mentioned before, results of \citet{Lessmann2015} or \citet{BROWN2012} also recommend the use of RF in the credit scoring context.
Parallel random forests (parRF) are a variation of RF that mainly decrease computational time by parallelizing the generation of random forests. 
However, they do not differ significantly in their performance to traditional RF \citep{Mitchell:2011}.

In contrast, rotation forests (rotForest) are an evolution from RF. 
rotForest split the feature space into subsets on which a principal component analysis is applied. 
Then, a new feature space is reassembled while keeping all components in order to ensure variability in the data. 
The data are linearly transformed into new variables and used to train a decision tree.
As the decision trees' splits of the feature set will vary, different rotations are obtained, resulting in a diversity of classifiers \citep{rodriguez2006}.

\subsubsection{Model Averaging}
The last homogeneous ensemble approach is model averaging. This study presents results for model averaged neural networks (avNNEt). Using neural network models based on \citet{ripley2007} as individual classifiers, the same base model is fit to the data using different random number seeds. Using all resulting models for prediction, the model scores are averaged and translated into classes.

\subsection{Heterogeneous ensembles}
Like homogeneous ensembles, heterogeneous ensemble methods pool the predictions of many classifiers. 
In contrast to the former, the latter do so by combining different types of base models as they may have different views on the data and therefore complement each other in order to improve predictions.

\subsubsection{Averages}
For the simple (AvgS) and the weighted average (AvgW), predictions of all individual models as well as all homogeneous ensembles are combined using a simple or a weighted average respectively. 
The weights for the weighted average were calculated according to the models' predictive accuracy in the cross-validated performance evaluation in the training set.

\subsubsection{Stacking}
Stacking \citep[Stack, see, for example,][]{Breiman1996Stack} is a heterogeneous ensemble method which combines individual models of different kinds in order to obtain final predictions. 
Individual models can be considered first-level learners, the combining model is called a second-level learner or meta-learner \citep{zhou2012}. 
In a first step, the first-level learners are trained using the original data. 
Then, the output of this first step is used to create a new training data set for the meta-learner while keeping the class labels from the original data. 
In a second step, the meta-learner is trained on that new data set and predicts the class on its basis. In this study, all individual models and homogeneous ensemble methods were used as first-level learners and a stochastic gradient boosting model was used as the meta-learner combining the information from the base models.

\section{Experimental design}
\label{sec:design}
\subsection{Data sets}

As mentioned before, four different data sets are used in this study to compare the predictive performance of the classification techniques described above. 
The first two are among the most used data sets in the credit scoring literature, namely \textit{German Credit} (GC) and \textit{Australian Credit} (AC) from the UCI  Machine Learning Repository \citep{Dua2017}. 
The third data set, \textit{Taiwanese Credit} (TC) from \citet{Yeh2009} is a data set containing information about customers' default payments in Taiwan and was accessed via \citep{Dua2017}. 
The last and fourth data set, \textit{Give me some credit} (GMSC), was provided by a financial institution for the 2011 Kaggle Competition \citep{Kaggle2011}. 
An overview over the data sets' characteristics is available in Table \ref{tab:tab_data}.
These data sets vary with respect to the number of observations, the number of predictive variables as well as the prior default rate, i.e., the share of defaulting customers in the sample. 
While the number of predictive variables and instances in a data set is compelling from a computational point of view, the varying default rates in the data are of bigger interest for this study. 
While the prior default rate in GC and TC are fairly high (30\% and 22\% of all instances defaulted in the respective data set), AC exhibits an unusually high prior default rate, namely 55\%. 
GMSC, on the other hand, has by far the lowest prior default rate, with only 7\% of the instances in the data defaulting.

\begin{landscape}
\begin{table*}[t]
\caption{Classification models considered in the benchmarking study}\vspace*{-1.5em}\footnotesize
\begin{center}
\begin{threeparttable}
	\begin{tabular*}{23cm}{@{\extracolsep{\fill}} lllllcccr}
		\toprule
		& Name & Short Name & Libraries & Parameter & Num. & Fac. & NAs & Models\\
        \midrule 
        \multirow{10}{*}{\shortstack{Individual \\ models}} & Linear discriminant analysis & LDA$^1$ & \texttt{MASS} & - & X & X & & 1 \\
         & Quadratic discriminant analysis & QDA$^1$ & \texttt{MASS} & - & X & X & & 1 \\
         & Flexible discriminant analysis & FDA$^2$ & \texttt{earth}, \texttt{mda} & degree, nprune & X & X & & 650 \\
         & Logistic Regression & LogReg & \texttt{-} & - & X & X & & 1\\
         & Linear Support Vector Machine & SVM-L$^3$ & \texttt{kernlab} & C & X & X & X & 90 \\
         & SVM with radial basis kernel function & SVM-R$^3$ & \texttt{kernlab} & sigma, C & X & X & & 630 \\
         & k-Nearest Neighbors & KNN$^4$ & \texttt{-} & k & X & & & 200 \\
         & Artificial Neural Networks & ANN$^5$ & \texttt{nnet} & size, decay & X & X & & 120 \\
         & CART Decision Trees & CART$^6$ & \texttt{rpart} & cp & X & X & X & 80 \\
         & Naive Bayes Classifier & NB & \texttt{klaR} & fL, usekernel, adjust & X & X & X & 40 \\
         \midrule
         \multirow{12}{*}{\shortstack{Homogeneous \\ ensembles}}& Bagged Classification Tree & Bag-CT$^7$ & \texttt{ipred}, \texttt{plyr}, \texttt{e1071} & - & X & X & X & 1 \\
         & Boosted Classification Tree & Boost-CT$^{8}$ & \texttt{party}, \texttt{mboost}, \texttt{plyr} & mstop, maxdepth & X & X & X & 3500 \\
         & AdaBoost & AdaBoost$^{9}$ & \texttt{adabag}, \texttt{dplyr} & nlter, method & X & X & & 750 \\
         & Stochastic Gradient Boosting & SGB$^{10}$ & \texttt{gbm}, \texttt{plyr} & n.trees, interaction.depth, & X & X & X & 11250 \\
         &  &  &  & shrinkage, n.minobsinnode &  &  &  &  \\
         & Random Forests & RF$^{11}$ & \texttt{randomForest} & mtry & X & X & & 210 \\
         & Parallel Random Forest & parRF$^{12}$ & \texttt{e1071}, \texttt{randomForest}, & mtry & X & X & & 210 \\
         &  &  & \texttt{foreach}, \texttt{Import} &  &  &  & &  \\
         & Rotation Forest & rotForest$^{13}$ & \texttt{rotForest} & K, L & X & X & & 270\\
         & Model Averaged Neural Network & avNNet$^{14}$ & \texttt{nnet} & size, decay, bag & X & X & & 240 \\
         & Boosted Logistic Regression & Boost-Logit$^{8}$ & \texttt{caTools} & nlter & X & X & & 100 \\
         & Bagged FDA & Bag-FDA$^{15}$ & \texttt{earth}, \texttt{mda} & degree, nprune & X & X & & 2250 \\
         \midrule
         \multirow{3}{*}{\shortstack{Heterogeneous \\ ensembles}} & Simple average ensemble & AvgS & - &  & X &  & & 1 \\
         & Weighted average ensemble & AvgW & - & & X & & & 1 \\
         & Stacking & Stack $^{16}$ & \texttt{caret} & & X & & & 3\\
        \bottomrule
	\end{tabular*}
\begin{tablenotes}[para,flushleft]
\footnotesize{\textit{Notes}: $^1$\citet{ripley2007}, $^2$\citet{hastie1994}, $^3$\citet{Cortes1995}, $^4$\citet{Altman1992}, $^5$\citet{Sarle1994}, $^6$\citet{cart84}, $^7$\citet{Breiman1996Bag}, $^8$\citet{Schapireetal2012}, $^{9}$\citet{freund1997}, $^{10}$\citet{friedman2001}, \citet{FRIEDMAN2002}, $^{11}$\citet{Breiman2001}, $^{13}$\citet{rodriguez2006}, $^{14}$\citet{ripley2007}, $^{15}$\citet{Friedman1991}, $^{16}$\citet{Breiman1996Stack}.}
\end{tablenotes}
\end{threeparttable}
\end{center}
\label{tab:tab_methods}
\end{table*}
\end{landscape}

As computational aspects are not the focus of this study and the number of models estimated is substantial, only a subset of the larger data sets, namely TC and GMSC, is used in all calculations.
10\% and 5\% of the data sets TC and GMSC are used, respectively. 
The subsets are randomly sampled from the original data while maintaining the original default rate in the subset, which is central in finding the best techniques for handling data with unequal class distributions. Supplementary material in \cref{app:additionalres} provides summary statistics of both the original and the reduced data and shows that the data structure does alter significantly.

\begin{table*}[ht]
\caption{Credit Scoring data sets}\vspace*{-1.5em}\small
\begin{center}
\begin{threeparttable}
\begin{tabular*}{\textwidth}{@{\extracolsep{\fill}} lllccl}
\toprule
Name & Observations & \# Subset & Predictors & Prior default rate & Source\\
\midrule
GC & 1000 & 1000 & 20 & 0.30 & \citet{Dua2017}\\
AC & 690 & 690 & 14 & 0.55 & \citet{Dua2017}\\
TC & 30000 & 3001 & 24 & 0.22 &  \citet{Yeh2009}\\
GMSC & 150000 & 7501 & 10 & 0.07 & \citet{Kaggle2011}\\
  \bottomrule
\end{tabular*}
\begin{tablenotes}[para,flushleft]
\footnotesize{\textit{Notes}: Data sets, GC -- German Credit, AC -- Australian Credit, TC -- Taiwanese Credit, GMSC -- Give me some credit.}
\end{tablenotes}
\end{threeparttable}
\end{center}
\label{tab:tab_data}
\end{table*}

\subsection{Data preprocessing and partitioning}
The step of data preprocessing is vital in order to ensure comparability of the models, as some of them require the data adopted to be in a certain format.
For example, as not all classification methods used can work with factorial input, consequently, categorial variables are transformed into dummy variables where one base category is omitted to ensure full rank of the data, which is another requirement for some models.

Furthermore, it is important to test whether there is enough variation in all predictors, as some classifiers, for example tree based models, become unstable if a variable only has a single unique value. 
Therefore, all predictors in the data sets are checked for (near) zero variance and problematic predictors are removed from the data.
Highly correlated predictors pose another challenge for some classification models. 
Thus, the pairwise correlation between the predictors of one data set is calculated and if two variables have an absolute correlation higher than 0.9, one of those variables, namely the one with the highest correlation to other predictors, is removed from the data set.

In a next step, the data are randomly split into two partitions. 
75\% of the data are allocated into the training set, which will be utilized to train and calibrate the used models. 
The remaining 25\% of the data form the test data which serve as a hold-out-sample in order to evaluate the models' performance. 
In the splitting process, the overall class distribution of the data sets is maintained so that the prior default rate remains roughly the same in both, the training and the test set.
After splitting the data, numerical predictors are re-scaled into the interval between zero and one and missing values in the data are imputed using a k-nearest neighbor method. 
The five closest neighbors of an observation with a missing value are found based on Euclidean distance and the value for the predictor is imputed using the mean of these values. 
However, it is worth noticing that there are no missing values in the target variables in the data in use.

After all these preprocessing steps have been conducted, different sampling methods are applied to all four data sets. 
Using up-sampling, down-sampling, SMOTE, BorderlineSMOTE and ROSE results in a total of 24 different data sets with varying prior default rates (DR). 
An overview is provided in Table \ref{tab:tab_sampling}.

As shown in Table \ref{tab:tab_methods}, many of the classification methods are based on the use of one or more parameters, which ensures the optimal performance of the respective model. 
Consequently, before the performance of all classifiers can be compared, the models in use have to be calibrated in order to find an optimal choice of parameters. 
This optimization is done using a grid of possible values for the parameters and 10-fold cross-validation within the training set: The training set is split into ten partitions or folds, and for each of the folds comprising 10\% of the training data, a model is built on the remaining 90\% of the data. 
The last 10\% are then used as reference to check the predictive performance of that model.
This implies that ten models are built on the training data and the average performance across all folds is reported. 
The process of 10-fold cross validation is conducted for each value and each parameter in the grid and the values, which result in the best model performance, are chosen for the final model. 
This strategy results in a multiplicity of models that are calculated and compared in order to find a best-performing model specification.

\subsection{Sampling methods}
As mentioned before, this study not only aims to evaluate the relative performance of classifiers in the face of class imbalances, but also identifies a best strategy in absolute terms in how to deal with unbalanced data. 
In order to do so, five sampling strategies are employed on all data sets and the classifiers in question are applied to them.

\begin{table*}[ht]
\caption{Sub-sampling methods in the training data, non-default/default cases and default ratios}\vspace*{-1.5em}\scriptsize
\begin{center}
\begin{threeparttable}
\begin{tabular*}{\textwidth}{@{\extracolsep{\fill}} lllllllllllllll}
	\toprule
     & \multicolumn{2}{c}{original} & \multicolumn{2}{c}{down-sampling} & \multicolumn{2}{c}{up-sampling} & \multicolumn{2}{c}{SMOTE} & \multicolumn{2}{c}{ROSE} & \multicolumn{2}{c}{BSMOTE}\\
    \cmidrule(l{3pt}r{3pt}){2-3}\cmidrule(l{3pt}r{3pt}){4-5}\cmidrule(l{3pt}r{3pt}){6-7}\cmidrule(l{3pt}r{3pt}){8-9}\cmidrule(l{3pt}r{3pt}){10-11}\cmidrule(l{3pt}r{3pt}){12-13}
    Data set & ND/D & DR & ND/D & DR & ND/D & DR & ND/D & DR & ND/D & DR & ND/D & DR \\
    \midrule
    GC & 525/225 & 0.30 & 225/225 & 0.50 & 525/525 & 0.50 & 900/675 & 0.43 & 381/369 & 0.49 & 525/225 &  0.30\\
    AC & 231/288 & 0.55 & 231/231 & 0.50 & 288/288 & 0.50 & 693/924 & 0.57 & 241/278 & 0.54 & 231/288 & 0.55\\
    TC & 1753/498 & 0.22 & 498/498 & 0.50 & 1753/1753 & 0.50 & 1992/1494 & 0.43 & 1135/1116 & 0.50 & 1753/498 & 0.22\\
    GMSC & 5250/377 & 0.07 & 144/144 & 0.50 & 5250/5250 & 0.50 & 1508/1131 & 0.43 & 2834/2793 & 0.50 & 5250/377 & 0.07\\
    \bottomrule
\end{tabular*}
\begin{tablenotes}[para,flushleft]
\footnotesize{\textit{Notes}: Non-default (ND), default (D) and default ratio (DR). Data sets: GC -- German Credit, AC -- Australian Credit, TC -- Taiwanese Credit, GMSC -- Give me some credit.}
\end{tablenotes}
\end{threeparttable}
\end{center}
\label{tab:tab_sampling}
\end{table*}

Sampling strategies alter the class distribution in the original training set in an additional preprocessing step, while the test data remain unchanged. 
A model is then trained with that data and evaluated on the (still unbalanced) test data.
Sampling strategies are not the only approach to classifying unbalanced data. 
One example are cost-sensitive learning methods which make use of a cost matrix which contains a class misjudged punishment coefficient in order to raise the misjudgment cost weight of default samples. 
This simplifies the original classification task into an optimization problem to minimize the misclassification of all observations. 
The study by \citet{hand2003} is an example that accounts for misclassification costs in the solution to a classification problem with imbalanced data, with relatively good results.

Another approach addresses the issue that sampling methods often produce redundant samples, which do not contribute to the classification model or can be replaced by other samples which can increase the computational burden significantly. 
\citet{liu2009} put forward a BalanceCascade approach, which is an informed under-sampling technique overcoming this weakness of lost information of sampling techniques by randomly removing redundant samples with random under-sampling techniques. 
It also removes correctly classified samples in each iteration of the algorithm, which are useless in subsequent classifications and only keeps instances that bear useful information. 
\citet{He2018} took this technique of combining classification models with sampling further. 
The authors construct a novel ensemble model to improve predictive performance in the face of different imbalance data ratios. 
Using an extended BalanceCascade approach, adjustable data subsets are built by estimating the data imbalance ratios. 
These subsets are then used for training tree-based base classifiers, which are then stacked together to an ensemble which produces predictions that are then used as covariates in another stacking layer.

While all strategies concerning imbalanced data have different foci, sampling methods have the advantage that they require no knowledge about misclassification costs and can be combined with any classification method, as the only change required is to the training data rather than to the model itself \citep{Drummond2003}. 
As a result, sampling methods prove to be an alternative that is flexible and relatively easy to implement as one can make use of the large variety of already implemented classification methods and combine them with different sampling strategies.

Under-sampling or down-sampling is a fairly simple approach which creates a new data set by randomly sampling from the original data so that all classes have the same frequency as the minority class.

Oversampling or up-sampling on the other hand increases the samples of the minority class by sampling with replacement from the original data, until the class distributions are equal. However, this approach increases the risk of overfitting the model in the training set, possibly resulting in poor predictive performance when the model is evaluated on the test data.

\citet{chawla2002} proposed the Synthetic Minority Oversampling Technique (SMOTE) to overcome this problem by blending under-sampling of the majority class with a special form of over-sampling the minority class. 
New samples are inserted between minority class instances and its neighbors instead of directly oversampling with replacement instances of the minority class. 
Each minority class observation is sampled and synthetic observations are introduced in the neighborhood of the $k$ minority class nearest neighbors. 
The new synthetic examples lead to a more general data space with respect to the minority class, so that the surrounding majority instances do not subsume the minority class instances. 
In addition to the creation of synthetic examples, the number of instances from the majority class, which is supposed to be sampled for each instance of the minority class, can also be specified so that the ratio between both classes reaches a specified point, effectively resulting in under-sampling of the majority class. 
\citet{chawla2002} tested the approach using C4.5 decision trees, Ripper and NB classifiers and concluded that SMOTE improved accuracy for those classifiers when compared to simple under-sampling.

However, \citet{He2018} point out that some weaknesses exist in the SMOTE approach. For example, the selected neighbor and the current instance need not be in the same class. 
As a result, improved versions of SMOTE are proposed, e.g., BorderlineSMOTE (BSMOTE) by \citet{han2005}. 
BSMOTE aims at improving classifications by only over-sampling the minority instances that lie near the borderline separating one class from the other. 
The instances on or nearby the borderline are more likely misclassified than those far away from it. 
Consequently, they bear more important information for the classification process.

Another sampling technique is the Random Over Sampling Examples approach (ROSE) proposed by \citet{Menardi2014}. 
It is a smoothed bootstrap-based technique building on the generation of new artificial examples from the classes. 
The algorithm essentially draws an observation from the training set belonging to one of the two classes, then, a new instance is generated in its neighborhood. 
While ROSE may be considered a special case of oversampling, in contrast to the latter, the risk of overfitting can be reduced following this approach.

\subsection{Performance measures}
Once the classifiers are applied to all variations of the original data sets, their predictive performance is evaluated by employing the following performance metrics.

Accuracy, or the percentage of correctly classified (PCC), is one of the simplest and most widely used performance measures in the classification literature. 
It is the fraction of correctly classified observations within a given test set and, therefore, requires discrete class predictions. 
Those can be retrieved by comparing $p(+|\textbf{x}_i)$ to the previously mentioned threshold $\tau$.
Given discrete class predictions, a confusion matrix (see e.g., Table \ref{tab:tab_cm}), which gathers information about the class predictions and actual classes in a test set, can be obtained.

\begin{table*}[ht]
\caption{Confusion matrix}\vspace*{-1.5em}
\begin{center}
\begin{threeparttable}
\begin{tabular*}{\textwidth}{@{\extracolsep{\fill}} ll|ll}
\toprule
    & & \multicolumn{2}{c}{Predicted} \\
    & & No Default (Positive) & Default (Negative)\\
    \midrule
    \multirow{2}{*}{Actual} &  \shortstack{No Default (Positive)} & True Positives (TP) & False Negatives (FN)\\
    & \shortstack{Default (Negatives)} & False Positives (FP) & True Negatives (TN)\\
    \bottomrule
\end{tabular*}
\end{threeparttable}
\end{center}
\label{tab:tab_cm}
\end{table*}

The accuracy measure is given by
\begin{equation}
    \text{Accuracy}=\frac{\text{TP} + \text{TN}}{\text{TP} + \text{FN} + \text{FP} + \text{TN}},
\end{equation}

where TP stands for True Positives, the ratio of actual positive instances which were classified as positive, FN stands for False Negatives, the fraction of actual positive cases which are labelled as negative. FP and TN denote False Positives, negative cases wrongly classified as positives, and True Negatives, correctly labelled negative cases, respectively. The accuracy metric ranges from zero to one, where the higher the metric, the more accurately a model can predict classes in a data set. 
While accuracy and error rate, which is given by $1-\text{Accuracy}$, are reasonable performance measures in balanced data sets, other metrics are more appropriate in the presence of unbalanced data with unequal error costs, as it is the case in credit scoring, where labelling a defaulting customer as a ``good'' borrower is more expensive than the other way around.

Cohen's $\kappa$ \citep{cohen1960} is also based on the confusion matrix, but adjusts accuracy by accounting for the possibility of a correct prediction by chance alone. 
While accuracy cannot distinguish between a correct prediction based on class frequencies in the data and a correct prediction based on a good classification model, $\kappa$ rewards classifiers only when they are correct more often than by always guessing the most frequent class. 
$\kappa$ ranges from zero to one, where the upper bound indicates perfect agreement between predictions and actual classes and zero indicates perfect disagreement between the two. 

$\kappa$ is defined as
\begin{equation}
    \kappa = \frac{p_a - p_e}{1 - p_e},
\end{equation}
where $p_a$ is the proportion of actual agreement, i.e., accuracy, and $p_e$ is the expected agreement between predictions and the true classes based on the data's distribution.

While accuracy and $\kappa$ assess the correctness of categorial predictions, the AUC, short for area under the curve, assesses a classifier's discriminatory ability. 
It is based on the Receiver Operating Characteristic (ROC) curve and represents the area under the ROC curve. 
In the ROC curve, the true positive rate (TPR), or Sensitivity, on the vertical axis is plotted against the false-positive rate (FPR), or $1-\text{Specifity}$, on the horizontal axis, where
\begin{align}
    \text{Sensitivity} &= \text{TPR} =  \frac{\text{TP}}{\text{TP} + \text{FN}},\label{eq:TPR}\\
    1-\text{Specificity} &= \text{FPR} = \frac{\text{FP}}{\text{TN} + \text{FP}}.\label{eq:FPR}
\end{align}
The AUC reduces this two-dimensional depiction of classification performance into a single scalar, which represents the probability that a classifier will rank a randomly chosen positive instance higher than a randomly chosen negative one \citep{FAWCETT2004}. Consequently the AUC ranges from 0.5 to 1, where a classifier with an AUC of 0.5 has no predictive value and a higher score implies a better predictive performance. 
One big deficiency of the AUC is that it assumes different cost distributions for different classification methods, whereas the distribution of misclassification costs actually depends on the classification problem at hand, and not the classification model in use.

The H-Measure, proposed by \citet{Hand2009}, is another performance metric which evaluates a classifier's discriminatory ability, like the AUC, but it can overcome the deficiency of different cost distributions of the latter. 
In order to do so, the relative classification costs are specified using a beta-distribution, which makes the H-Measure consistent across classifiers.

Covering another aspect of predictive performance, the Brier Score (BS) evaluates the accuracy of probability predictions. 
It is equal to the mean squared error of a probability estimate $p(+|\textbf{x}_i)$ and a zero and one response variable.

The last performance metric, the Kolmogorov-Smirnov statistic (KS), assesses a model's capability to identify positive and negative instances in the data correctly \citep[see, e.g.,][]{He2018}. 
It is given by the maximum difference between the cumulative score distribution of positive and negative, i.e.,
\begin{equation}
    KS = \underset{q}{\max}(|\text{TPR}(q) - \text{FPR}(q)|),
\end{equation}
where $q$ stands for the cumulative quantile and TPR$(q)$ and FPR$(q)$ represent the values of TPR and FPR, respectively, when the cumulative quantile is accumulated to $q$. 
TPR and FPR are defined as in equations \ref{eq:TPR} and \ref{eq:FPR}, respectively.

A larger value of the KS implies better predictive performance of a classifier.

All these performance metrics are flawed in some aspect for the given credit scoring classification problem. 
\citet{Lessmann2015} summarize some of the disadvantages of the given metrics. 
As mentioned before, accuracy is considered a dismal performance indicator for imbalanced data. Furthermore, measures that evaluate categorial predictions such as accuracy, $\kappa$ or KS depend on the choice of $\tau$, and different choices of the threshold can result in misleading values for accuracy and $\kappa$. 
AUC and BS are global performance indicators, as they consider the whole score distribution. 
However, such a perspective implies that all thresholds are equally probable, an assumption that is not plausible in credit scoring \citep{Hand2005}. 
As only customers or loan applicants with a probability above the threshold will be accepted, accuracy is particularly important in the upper tail of the distribution.
Using several performance measures together and summarizing a classifier's rank over all of them should lead to a reliable estimate of its predictive performance. 
Therefore, this study reports the classifiers' ranks in all six performance measures described above as well as the average rank across all measures.

\section{Performance evaluation}\label{sec:results}
The following section presents the benchmarking results of the 23 classifiers across four credit scoring data sets and for each sampling strategy. A model's predictive performance is evaluated across the six different performance measures described above. For each data set and performance indicator, a classifier's rank was evaluated. Tables \ref{table:results1} to \ref{table:results6} present the average rank of all classifiers across the four data sets for each performance metric as well as the average rank (AvgR) of each classifier across all measures. Based on AvgR, an overall rank of the classifier is included in the last column of the result tables. All performance measures for all data sets are included in the supplementary materials provided in \cref{app:additionalres}.

\begin{table*}[t]
\caption{Results original data sets}\vspace*{-1.5em}\small
\begin{center}
\begin{threeparttable}
\begin{tabular*}{\textwidth}{@{\extracolsep{\fill}} lllllllll}
\toprule
 & PCC & $\kappa$ & BS & AUC & HM & KS & AvgR  & Rank\\ 
  \hline
  LDA & 14.1 & 13.8 & 13.9 & 16.0 & 14.8 & 13.9 & 14.4 & 16 \\ 
  QDA & 17.9 & 17.6 & 17.6 & 18.0 & 19.3 & 17.3 & 17.9 & 23 \\ 
  FDA & \textbf{5.4} & \textbf{4.9} & \textbf{5.1} & 12.3 & 12.3 & 12.8 & 8.8 & 9 \\ 
  LogReg & 13.9 & 13.5 & 13.6 & 17.3 & 16.5 & 16.9 & 15.3 & 18 \\ 
  SVM-L & 15.4 & 15.0 & 15.1 & 15.3 & 13.5 & 14.6 & 14.8 & 17 \\ 
  SVM-R & 18.8 & 18.5 & 18.5 & 16.3 & 13.0 & 13.0 & 16.3 & 20 \\ 
  KNN & 12.5 & 13.0 & 12.3 & 18.3 & 20.3 & \underline{20.8} & 16.2 & 19 \\ 
  ANN & 11.6 & 11.1 & 11.4 & 10.5 & 11.3 & 10.4 & 11.0 & 11 \\ 
  CART & 6.6 & 6.6 & 6.4 & \underline{21.8} & 20.8 & \underline{21.3} & 13.9 & 14 \\ 
  NB & 16.5 & 16.0 & 16.3 & 17.5 & 17.5 & 16.4 & 16.7 & 21 \\ 
  \hline
  Bag-CT & 15.1 & 14.5 & 14.9 & 11.9 & 14.0 & 12.1 & 13.8 & 13 \\ 
  Boost-CT & 8.1 & 7.4 & 7.9 & \textbf{3.0} & 4.5 & 5.9 & \textbf{6.1} & 1 \\ 
  AdaBoost & 10.5 & 10.0 & 10.3 & 3.6 & 4.5 & \textbf{3.3}& 7.0 & 2 \\ 
  SGB & 9.4 & 9.1 & 9.1 & 5.3 & 5.3 & 5.9 & 7.3 & 3 \\ 
  RF & 9.9 & 9.3 & 9.6 & 6.8 & 7.0 & 8.3 & 8.5 & 5 \\ 
  parRF & 9.5 & 9.5 & 9.3 & 8.0 & 6.8 & 8.1 & 8.5 & 6 \\ 
  rotForest & 11.4 & 16.6 & 16.9 & 19.3 & 19.0 & 18.8 & 17.0 & 22 \\ 
  avNNet & 15.5 & 15.3 & 15.3 & 14.0 & 12.8 & 12.9 & 14.3 & 15 \\ 
  Logit-Boost & 7.6 & 9.8 & 7.4 & 18.3 & 19.3 & 19.8 & 13.7 & 12 \\ 
  bagFDA & 8.1 & 7.6 & 7.9 & 8.5 & 9.5 & 8.4 & 8.3 & 4 \\ 
  \hline
  AvgS & 13.4 & 13.1 & 13.1 & 4.6 & \textbf{4.3} & 3.8 & 8.7 & 8 \\ 
  AvgW & 13.4 & 13.1 & 13.1 & 3.9 & 4.3 & 4.0 & 8.6 & 7 \\ 
  Stack & 11.5 & 10.8 & 11.3 & 6.0 & 6.0 & 7.9 & 8.9 & 10 \\
  \hline
  Friedman $\chi^2_F$ & 25.868  & 26.844  & 28.059 & \underline{67.794} & \underline{64.212} & \underline{60.700} &  & \\
  & \small{(0.257)} & \small{(0.217)} & \small{(0.174)} & \small{(0.000)} & \small{(0.000)} & \small{(0.000)} & & \\
  \bottomrule
\end{tabular*}
\begin{tablenotes}[para,flushleft]
\footnotesize{\textit{Notes}: Bold face indicates the best classifier (i.e., with the lowest average rank over the data sets) per performance metric. The last row shows the Friedman statistic with p-values. Underlined values indicate that a classifier performs significantly worse than the best classifier in a data scenario at the 5 percent level according to the Nemenyi test or, in the case of the Friedman statistic, indicate that not all classifiers perform equally well.}
\end{tablenotes}
\end{threeparttable}
\end{center}
\label{table:results1}
\end{table*}

This study adopts the nonparametric testing framework proposed by \citet{demvsar2006} for comparing classifiers over different data sets. 
In a first step, the Friedman test \citep{friedman1937use, friedman1940comparison}, a non-parametric equivalent of the repeated-measures ANOVA, is conducted for each performance measure. The Friedman statistic $\chi_F^2$ and its p-values (in brackets) are reported at the bottom of each result table for all performance measures, respectively. 
The null-hypothesis, stating that all models perform equally well, is rejected if $p<0.05$, indicated by an underlined Friedman statistic.

Once the null-hypothesis is rejected, the Nemenyi test is performed as a post-hoc test. 
The Nemenyi test is used to compare classifiers pairwise, where the best performing classifier per measure is tested against all other models. 
If a classifier performs significantly worse than the best model, its rank is underlined in all tables.
However, for the vast majority, the results of this test point to no or very few significant differences in the models' performances.
In the following, the benchmarking results will be discussed in more detail for each data sampling scenario before drawing overall conclusions.

The performance metrics for all classifiers on the original data sets, which have undergone no sampling procedures, are provided in Table \ref{table:results1}. 
The overall best-performing classifier on the original credit scoring data is Boost-CT, the boosted classification tree. 
While the Friedman statistics for accuracy, $\kappa$ and the Brier score are not significant and thus we cannot assume that the classifiers differ in terms of these three metrics, two classifiers, namely CART and KNN, perform significantly worse than Boost-CT according to two other test metrics. 
CART's predictions are less accurate in terms of AUC and KS, KNN predicts worse in terms of KS. 

Interestingly, FDA performs relatively well in terms of PCC, $\kappa$ or BS but quite poorly in terms of AUC, HM and KS, resulting in a low overall rank, the same holds for CART. At least for those two classifiers, this seems to confirm the importance of using several different performance measures in order to obtain a reliable impression of the quality of a model's predictions, as different metrics may give contradictory results. 

\begin{table*}[t]
\caption{Results down-sampled data sets}  \vspace*{-1.5em}\small
\begin{center}
\begin{threeparttable}
\begin{tabular*}{\textwidth}{@{\extracolsep{\fill}} lllllllll}
\toprule
 & PCC & $\kappa$ & BS & AUC & HM & KS & AvgR  & Rank\\  
  \midrule
  LDA & 17.0 & 17.1 & 16.0 & 17.3 & 16.8 & 18.4 & 17.1 & 20 \\ 
  QDA & 19.0 & 18.8 & 18.3 & 18.0 & 18.8 & 18.0 & 18.5 & 22 \\ 
  FDA & 12.8 & 15.8 & 16.3 & 14.0 & 14.0 & 13.5 & 14.4 & 15 \\ 
  LogReg & 16.3 & 16.4 & 15.3 & 17.3 & 15.0 & 15.8 & 16.0 & 18 \\ 
  SVM-L & 18.4 & 18.3 & 17.4 & 16.5 & 15.5 & 18.3 & 17.4 & 21 \\ 
  SVM-R & 10.4 & 10.3 & 9.6 & 13.3 & 9.3 & 12.0 & 10.8 & 11 \\ 
  KNN & 13.1 & 14.8 & 12.1 & 18.8 & 20.3 & 19.9 & 16.5 & 19 \\ 
  ANN & 16.1 & 16.0 & 15.4 & 15.3 & 15.0 & 14.1 & 15.3 & 17 \\ 
  CART & 6.3 & 7.1 & 6.3 & 17.1 & 18.4 & 15.9 & 11.8 & 13 \\ 
  NB & 19.0 & 18.8 & 18 & 19.0 & 18.8 & 18.5 & 18.7 & 23 \\ 
  \midrule
  Bag-CT & 7.6 & 9.8 & 7.6 & 12.0 & 13.3 & 14.3 & 10.8 & 10 \\ 
  Boost-CT & 7.5 & 8.0 & 7.3 & 5.3 & 5.0 & 7.0 & 6.7 & 2 \\ 
  AdaBoost & 9.0 & 8.4 & 8.3 & \textbf{2.8} & 6.0 & 7.0 & 6.9 & 4 \\ 
  SGB & 12.4 & 11.8 & 12.4 & 4.3 & 5.0 & 5.5 & 8.5 & 8 \\ 
  RF & 8.3 & 7.1 & 8.0 & 6.5 & 6.0 & 5.4 & 6.9 & 3 \\ 
  parRF & \textbf{6.0} & \textbf{6.1} & \textbf{6.0} & 6.3 & \textbf{3.3} & 6.3 & \textbf{5.6} & 1 \\ 
  rotForest & 7.1 & 6.8 & 6.4 & 16.4 & 16.4 & 15.1 & 11.4 & 12 \\ 
  avNNet & 13.9 & 14.0 & 12.9 & 14.8 & 13.5 & 11.8 & 13.5 & 14 \\ 
  Logit-Boost & 10.5 & 9.8 & 10.5 & 18.8 & 20.5 & 19.0 & 14.8 & 16 \\ 
  bagFDA & 9.3 & 8.4 & 8.5 & 5.8 & 6.8 & 8.8 & 7.9 & 5 \\ 
  \midrule
  AvgS & 11.9 & 10.5 & 14.8 & 4.3 & 5.3 & \textbf{3.5} & 8.4 & 7 \\ 
  AvgW & 11.9 & 10.5 & 15.0 & 3.3 & 3.5 & 4.0 & 8.0 & 6 \\ 
  Stack & 12.5 & 11.9 & 14.0 & 9.5 & 10.0 & 4.3 & 10.4 & 9 \\ 
  \midrule
  Friedman $\chi^2_F$ & 33.339  & \underline{34.385}  & 27.203 & \underline{65.837} & \underline{65.690} & \underline{60.563} &  & \\
   & \small{(0.057)} & \small{(0.045)} & \small{(0.204)} & \small{(0.000)}  & \small{(0.000)}  & \small{(0.000)}\\
  \bottomrule
\end{tabular*}
\begin{tablenotes}[para,flushleft]
\footnotesize{\textit{Notes}: Bold face indicates the best classifier (i.e., with the lowest average rank over the data sets) per performance metric. The last row shows the Friedman statistic with p-values. Underlined values indicate that a classifier performs significantly worse than the best classifier in a data scenario at the 5 percent level according to the Nemenyi test or, in the case of the Friedman statistic, indicate that not all classifiers perform equally well.}
\end{tablenotes}
\end{threeparttable}
\end{center}
\label{table:results2}
\end{table*}

In the data sets that have been preprocessed using down-sampling, Boost-CT performs relatively well again, but is outperformed by parRF, parallel random forest. 
Again, individual classifiers rank clearly below ensemble methods, although no model performs significantly worse than parRF according to the Nemenyi test. 
While boosting models perform relatively well again, parRF and RF now also rank amongst the best-performing classification techniques. 
This implies that these techniques still work relatively well with relatively small data sets, even though a lot of information is lost during the down-sampling process. 
In addition, heterogeneous ensembles rank slightly better when performed on down-sampled data than on the original data. 
Interestingly, both Table \ref{table:results1} and \ref{table:results2} also support the finding of \citet{Baesens2003} in that linear classifiers do not perform worse than non-linear ones, indicating that there is at most weak non-linearity in the data used.

\begin{table*}[t]
\caption{Results up-sampled data sets}  \vspace*{-1.5em}\small
\begin{center}
\begin{threeparttable}
\begin{tabular*}{\textwidth}{@{\extracolsep{\fill}} lllllllll}
\toprule
 & PCC & $\kappa$ & BS & AUC & HM & KS & AvgR  & Rank\\   
  \midrule
  LDA & 19.1 & 18.6 & 19.1 & 18.8 & 16 & 16 & 17.9 & 21 \\ 
  QDA & 18.0 & 17.5 & 18.0 & 19.0 & 19.5 & 18.8 & 18.5 & 22 \\ 
  FDA & 12.1 & 11.6 & 12.1 & 6.0 & 9.5 & 12.0 & 10.6 & 9 \\ 
  LogReg & 18.6 & 18.1 & 18.6 & 17.0 & 16.5 & 16.8 & 17.6 & 20 \\ 
  SVM-L & 17.0 & 16.5 & 17.0 & 16.3 & 13.8 & 13.5 & 15.7 & 18 \\ 
  SVM-R & 12.1 & 11.4 & 12.1 & 16.3 & 10.3 & 10.5 & 12.1 & 13 \\ 
  KNN & \textbf{2.5} & 13.0 & \textbf{2.5} & 19.0 & 19.5 & 17.8 & 12.4 & 14 \\ 
  ANN & 16.8 & 15.8 & 16.8 & 17.0 & 16.5 & 18.3 & 16.8 & 19 \\ 
  CART & 8.0 & 7.3 & 8.0 & 16.8 & 17.8 & 14.3 & 12.0 & 12 \\ 
  NB & \underline{21.5} & 21.0 & \underline{21.5} & 19.0 & 19.3 & 18.8 & 20.2 & 23 \\
  \midrule
  Bag-CT & 8.5 & 11.3 & 8.5 & 12.0 & 10.0 & 8.8 & 9.8 & 7 \\ 
  Boost-CT & 8.3 & 8.0 & 8.3 & 13.8 & 9.0 & 13.5 & 10.1 & 8 \\ 
  AdaBoost & 16.5 & 15.6 & 16.5 & 10.0 & 10.8 & 12.0 & 13.6 & 15 \\ 
  SGB & 5.8 & 5.0 & 5.8 & 4.3 & 7.5 & 8.3 & 6.1 & 4 \\ 
  RF & 7.8 & 7.0 & 7.8 & 5.8 & \textbf{3.3} & 6.8 & 6.4 & 5 \\ 
  parRF & 4.8 & \textbf{4.0} & 4.8 & 4.8 & 5.0 & 6.8 & \textbf{5.0} & 1 \\ 
  rotForest & 8.9 & 8.4 & 8.9 & 5.8 & 5.8 & 4.5 & 7.0 & 6 \\ 
  avNNet & 16.1 & 15.4 & 16.1 & 12.5 & 14.5 & 16.3 & 15.1 & 17 \\ 
  Logit-Boost & 12.5 & 12.0 & 12.5 & 16.0 & 19.0 & 16.8 & 14.8 & 16 \\ 
  bagFDA & 14.5 & 13.6 & 14.5 & 5.8 & 8.0 & 11.0 & 11.2 & 10 \\ 
  \midrule
  AvgS & 7.5 & 6.8 & 7.5 & 3.5 & 6.0 & \textbf{3.5} & 5.8 & 2 \\ 
  AvgW & 7.5 & 6.8 & 7.5 & \textbf{3.3} & 5.5 & 4.3 & 5.8 & 2 \\ 
  Stack & 11.8 & 11.5 & 11.8 & 13.8 & 13.3 & 7.3 & 11.5 & 11 \\ 
  \midrule
  Friedman $\chi^2_F$ & \underline{54.250} & \underline{44.936} & \underline{54.250} & \underline{64.559} & \underline{53.696} & \underline{47.536} &  &  \\ 
  & \small{(0.000)} & \small{(0.003)} & \small{(0.000)} & \small{(0.000)} & \small{(0.000)}& \small{(0.001)} & & \\
 \bottomrule
\end{tabular*}
\begin{tablenotes}[para,flushleft]
\footnotesize{\textit{Notes}: Bold face indicates the best classifier (i.e., with the lowest average rank over the data sets) per performance metric. The last row shows the Friedman statistic with p-values. Underlined values indicate that a classifier performs significantly worse than the best classifier in a data scenario at the 5 percent level according to the Nemenyi test or, in the case of the Friedman statistic, indicate that not all classifiers perform equally well.}
\end{tablenotes}
\end{threeparttable}
\end{center}
\label{table:results3}
\end{table*}

When applying the classifiers to the up-sampled data (see Table \ref{table:results3}), the Friedman statistic indicates that there are significant differences in the predictive performance of the models for all metrics. 
However, conducting the Nemenyi test only allows concluding that NB makes significantly worse predictions than the best algorithm, parRF. 
The models of the family of random forest classifiers (RF, parRF and rotForest) and boosting classifiers (Boost-CT, AdaBoost, SGB, LogitBoost), again, belong to the best performing classifiers. 
However, with the up-sampled data, bagging techniques like Bag-CT or bagFDA as well as simple and weighted averages, AvgS and AvgW, also rank quite well.

Table \ref{table:results4} provides the results for classifiers applied to data which were sampled using SMOTE. 
Again, the Friedman statistic leads to the rejection of the hypothesis that there are no differences between the predictive performance of the classifiers for all performance measures. 
According to the Nemenyi test, SVM-R and KNN both perform significantly worse in terms of AUC, HM and KS than the best algorithm, which is again Boost-CT. 
The next best models are, again, RF and parRF.

\begin{table*}[t]
\caption{Results data sets sampled using SMOTE}  \vspace*{-1.5em}\small
\begin{center}
\begin{threeparttable}
\begin{tabular*}{\textwidth}{@{\extracolsep{\fill}} lllllllll}
\toprule
 & PCC & $\kappa$ & BS & AUC & HM & KS & AvgR  & Rank\\
  \midrule
  LDA & 17.5 & 17.3 & 17.3 & 14.0 & 9.5 & 10.9 & 14.4 & 15 \\ 
  QDA & 17.3 & 16.3 & 17.0 & 16.8 & 16.8 & 16.0 & 16.7 & 21 \\ 
  FDA & \textbf{5.1} & \textbf{4.4} & \textbf{4.9} & 9.8 & 9.5 & 9.3 & 7.1 & 5 \\ 
  LogReg & 16.8 & 16.5 & 16.5 & 15.5 & 12.0 & 13.1 & 15.1 & 17 \\ 
  SVM-L & 19.0 & 18.8 & 18.8 & 15.3 & 12.0 & 10.9 & 15.8 & 18 \\ 
  SVM-R & 20.5 & 20.0 & 20.3 & \underline{20.5} & \underline{21.0} & \underline{21.5} & 20.6 & 23 \\ 
  KNN & 12.6 & 17.3 & 12.3 & \underline{22.5} & \underline{22.5} & \underline{22.5} & 18.3 & 22 \\ 
  ANN & 15.8 & 15.4 & 15.5 & 18.5 & 17.3 & 16.9 & 16.5 & 20 \\ 
  CART & 7.6 & 8.3 & 7.5 & 18 & 16.5 & 13.5 & 11.9 & 12 \\ 
  NB & 16.4 & 15.6 & 16.1 & 16.3 & 15.5 & 17.3 & 16.2 & 19 \\ 
  \midrule
  Bag-CT & 11.3 & 11.3 & 11.0 & 12.0 & 15.5 & 15.5 & 12.8 & 13 \\ 
  Boost-CT & 8.4 & 7.9 & 8.1 & 5.3 & 5.0 & 4.3 & \textbf{6.5} & 1 \\ 
  AdaBoost & 13.3 & 12.4 & 13 & 9.5 & 12.8 & 9.0 & 11.6 & 11 \\ 
  SGB & 6.8 & 6.0 & 6.5 & 6.4 & 7.3 & 8.8 & 6.9 & 4 \\ 
  RF & 7.0 & 6.3 & 6.8 & 4.8 & 7.3 & 7.1 & 6.5 & 2 \\ 
  parRF & 6.5 & 5.8 & 6.3 & 5.6 & 7.3 & 8.5 & 6.6 & 3 \\ 
  rotForest & 5.9 & 11.1 & 11.4 & 10.5 & 12.5 & 13 & 10.7 & 9 \\ 
  avNNet & 12.4 & 11.6 & 12.1 & 10.0 & 10.5 & 11.8 & 11.4 & 10 \\ 
  Logit-Boost & 10.5 & 11.0 & 10.3 & 18.3 & 20.0 & 19.1 & 14.9 & 16 \\ 
  bagFDA & 8.9 & 8.1 & 8.6 & 8.0 & 6.5 & 7.0 & 7.9 & 8 \\ 
  \midrule
  AvgS & 11.6 & 11.3 & 11.4 & 4.0 & \textbf{2.8} & \textbf{2.4} & 7.2 & 6 \\ 
  AvgW & 11.6 & 11.3 & 11.4 & \textbf{3.0} & 3.5 & 3.8 & 7.4 & 7 \\ 
  Stack & 13.5 & 12.5 & 13.3 & 11.8 & 12.8 & 14.1 & 13.0 & 14 \\ 
  \midrule
  Friedman $\chi^2_F$ & \underline{39.877} & \underline{39.066} & \underline{36.34}  & \underline{62.293}  & \underline{58.348} & \underline{56.676} &  & \\ 
  & \small{(0.011)} & \small{(0.011)} & \small{(0.014)} & \small{(0.028)} & \small{(0.000)} & \small{(0.000)} & &\\
 \bottomrule
\end{tabular*}
\begin{tablenotes}[para,flushleft]
\footnotesize{\textit{Notes}: Bold face indicates the best classifier (i.e., with the lowest average rank over the data sets) per performance metric. The last row shows the Friedman statistic with p-values. Underlined values indicate that a classifier performs significantly worse than the best classifier in a data scenario at the 5 percent level according to the Nemenyi test or, in the case of the Friedman statistic, indicate that not all classifiers perform equally well.}
\end{tablenotes}
\end{threeparttable}
\end{center}
\label{table:results4}
\end{table*}

This pattern changes when looking at Table \ref{table:results5}, which provides results for the data sets sampled using ROSE. 
Here, three individual classifiers, ANN, NB and SVM-L, rank amongst the ten best models. 

avNNet, the overall best ranked model, does not seem to be significantly better than any other method, but for once, it outperforms the classifiers that performed well in the other data scenarios. The two averages, AvgS and AvgW, again rank quite high overall.

\begin{table*}[t]
\caption{Results data sets sampled using ROSE} \vspace*{-1.5em}\small
\begin{center}
\begin{threeparttable}
\begin{tabular*}{\textwidth}{@{\extracolsep{\fill}} lllllllll}
\toprule
 & PCC & $\kappa$ & BS & AUC & HM & KS & AvgR  & Rank\\ 
  \midrule
  LDA & 16.4 & 14.5 & 15.9 & 12.5 & 7.8 & 7.9 & 12.5 & 14 \\ 
  QDA & 15.6 & 13.8 & 15.1 & 15.3 & 14 & 13.5 & 14.5 & 17 \\ 
  FDA & 18.3 & 17.5 & 17.8 & 16.8 & 16.5 & 14.3 & 16.8 & 22 \\ 
  LogReg & 14.4 & 12.6 & 13.9 & 11.5 & 8.0 & 6.4 & 11.1 & 11 \\ 
  SVM-L & 12.6 & 11.1 & 12.1 & 11.0 & 6.8 & 5.8 & 9.9 & 7 \\ 
  SVM-R & 15.5 & 13.3 & 15.0 & 10.8 & 12.0 & 12.6 & 13.2 & 15 \\ 
  KNN & 14.8 & 18.5 & 14.3 & 16.0 & 18.5 & 20.3 & 17.0 & 23 \\ 
  ANN & 9.8 & 7.8 & 9.3 & \textbf{5.0} & 5.3 & \textbf{5.3} & 7.0 & 2 \\ 
  CART & \textbf{4.5} & 12.9 & 9.8 & 22.3 & 22.0 & 20.8 & 15.4 & 20 \\ 
  NB & 11.4 & 10.0 & 10.9 & 9.3 & 10.5 & 6.0 & 9.7 & 6 \\ 
  \midrule
  Bag-CT & 7.6 & 12.3 & 12.9 & 13.5 & 18.3 & 18.5 & 13.8 & 16 \\ 
  Boost-CT & 13.4 & 11.9 & 12.9 & 16.8 & 17.8 & 19.3 & 15.3 & 19 \\ 
  AdaBoost & 18.0 & 18.5 & 17.5 & 15.0 & 14.5 & 16.0 & 16.6 & 21 \\ 
  SGB & 10.5 & 9.0 & 10.0 & 7.8 & 7.5 & 9.9 & 9.1 & 5 \\ 
  RF & 7.6 & 9.1 & 7.1 & 10.3 & 12.8 & 12.8 & 9.9 & 8 \\ 
  parRF & 7.1 & 10.3 & \textbf{6.6} & 11.8 & 13.5 & 14.0 & 10.5 & 9 \\ 
  rotForest & 9.5 & 11.8 & 9.0 & 15.0 & 13.0 & 15.8 & 12.3 & 13 \\ 
  avNNet & 8.9 & \textbf{7.4} & 8.4 & 5.0 & \textbf{5.0} & 5.9 & \textbf{6.8} & 1 \\ 
  Logit-Boost & 10.5 & 11.8 & 10.0 & 18.0 & 18.5 & 18.5 & 14.5 & 17 \\ 
  bagFDA & 12.9 & 11.5 & 12.4 & 11.5 & 9.8 & 8.3 & 11.0 & 10 \\ 
  \midrule
  AvgS & 11.4 & 9.3 & 10.9 & 5.8 & 5.5 & 7.5 & 8.4 & 3 \\ 
  AvgW & 11.4 & 9.3 & 10.9 & 6.8 & 6.0 & 8.0 & 8.7 & 4 \\ 
  Stack & 14.1 & 12.3 & 13.6 & 8.8 & 12.8 & 9.1 & 11.8 & 12 \\ 
  \midrule
  Friedman $\chi^2_F$ & 27.969  & 20.189  & 20.503  & \underline{38.283}  & \underline{48.13}  & \underline{51.689 } &  &  \\
  & \small{(0.177)} & \small{(0.571)} & \small{(0.552)} & \small{(0.017)} & \small{(0.001)} & \small{(0.000)} & & \\
 \bottomrule
\end{tabular*}
\begin{tablenotes}[para,flushleft]
\footnotesize{\textit{Notes}: Bold face indicates the best classifier (i.e., with the lowest average rank over the data sets) per performance metric. The last row shows the Friedman statistic with p-values. Underlined values indicate that a classifier performs significantly worse than the best classifier in a data scenario at the 5 percent level according to the Nemenyi test or, in the case of the Friedman statistic, indicate that not all classifiers perform equally well.}
\end{tablenotes}
\end{threeparttable}
\end{center}
\label{table:results5}
\end{table*}

Table \ref{table:results6}, providing the results for the data sampled using BSMOTE, offers a very similar picture as the results before. 
Again, the Friedman statistic indicates the existence of differences between the predictive performance between the classifiers. 
The Nemenyi test identifies KNN and Logit-Boost to perform worse than the best model, RF, in terms of AUC and HM, respectively. 
The best ten classifiers again belong to the group of ensemble methods, and within that group, RF, parRF, AvgS and AvgW rank best.

Generally, all these findings are mostly in line with the existing literature. 
They confirm the results of \citet{Nanni2009} and \citet{Lessmann2015}, who concluded that ensemble methods exceed the predictive performance of their base models or individual classifiers in general.

However, some conclusions contradict previous findings in the literature. 
ANN, which is very often cited as one of the most successful classification strategies in studies only comparing individual classifiers \citep[see, for example,][]{Yeh2009, Baesens2003} only performs well on data sampled with ROSE and adequately on the original data. 
In all other data scenarios, it is never the best individual classifier and ranks amongst the worst performing classifiers overall.

\begin{table*}[t]
\caption{Results data sets sampled using BSMOTE}\vspace*{-1.5em}\small
\begin{center}
\begin{threeparttable}
\begin{tabular*}{\textwidth}{@{\extracolsep{\fill}} lllllllll}
\toprule
 & PCC & $\kappa$ & BS & AUC & HM & KS & AvgR  & Rank\\ 
  \midrule
  LDA & 16.6 & 16.5 & 16.6 & 15.3 & 12.8 & 11.8 & 14.9 & 16 \\ 
  QDA & 17.1 & 17.0 & 17.1 & 18.8 & 17.3 & 17.3 & 17.4 & 21 \\ 
  FDA & 16.5 & 16.4 & 16.5 & 14.0 & 12.3 & 12.9 & 14.8 & 15 \\ 
  LogReg & 16.4 & 15.9 & 16.4 & 15.3 & 14.0 & 15.5 & 15.6 & 18 \\ 
  SVM-L & 15.5 & 15.4 & 15.5 & 15.8 & 13.0 & 14.5 & 14.9 & 17 \\ 
  SVM-R & 19.4 & 19.1 & 19.4 & 18.0 & 18.5 & 19.4 & 19.0 & 23 \\ 
  KNN & 11.6 & 16.1 & 11.6 & \underline{20.3} & 18.5 & 17.8 & 16.0 & 20 \\ 
  ANN & 15.5 & 15.1 & 15.5 & 15.5 & 16.0 & 16.6 & 15.7 & 19 \\ 
  CART & 7.9 & 6.8 & 7.9 & 16.3 & 17.0 & 13.9 & 11.6 & 11 \\ 
  NB & 19.1 & 18.9 & 19.1 & 18 & 18.5 & 17.8 & 18.6 & 22 \\ 
  \midrule
  Bag-CT & 8.5 & 9.6 & 8.5 & 11.5 & 13.8 & 12.8 & 10.8 & 9 \\ 
  Boost-CT & 11.6 & 11.0 & 11.6 & 9.8 & 7.5 & 9.0 & 10.1 & 7 \\ 
  AdaBoost & 12.5 & 12.0 & 12.5 & 12.3 & 13.3 & 14.8 & 12.9 & 13 \\ 
  SGB & 11.3 & 10.6 & 11.3 & 10.3 & 11.3 & 11.5 & 11.0 & 10 \\ 
  RF & \textbf{4.0} & \textbf{3.5} & \textbf{4.0} & 2.8 & \textbf{2.3} & \textbf{2.8} & \textbf{3.2} & 1 \\ 
  parRF & 9.4 & 9.0 & 9.4 & \textbf{2.3} & 4.8 & 5.3 & 6.7 & 4 \\ 
  rotForest & 9.8 & 9.5 & 9.8 & 10.0 & 12.5 & 10.9 & 10.4 & 8 \\ 
  avNNet & 11.8 & 10.8 & 11.8 & 11.8 & 11.8 & 13.1 & 11.8 & 12 \\ 
  Logit-Boost & 6.5 & 10.0 & 6.5 & 16.5 & \underline{20.0} & 17.8 & 12.9 & 13 \\ 
  bagFDA & 12.4 & 11.9 & 12.4 & 6.8 & 4.0 & 5.8 & 8.9 & 6 \\ 
  \midrule
  AvgS & 8.0 & 7.5 & 8.0 & 4.0 & 3.3 & 4.8 & 5.9 & 3 \\ 
  AvgW & 7.0 & 6.5 & 7.0 & 3.8 & 3.8 & 5.0 & 5.5 & 2 \\ 
  Stack & 7.8 & 7.0 & 7.8 & 7.5 & 10.3 & 5.5 & 7.6 & 5 \\ 
  \midrule
  Friedman $\chi^2_F$ & \underline{36.949}  & \underline{37.505}  & \underline{36.949}  & \underline{56.380} & \underline{56.522}  & \underline{48.663}  &  & \\ 
  & \small{(0.024)} & \small{(0.021)} & \small{(0.024)} & \small{(0.000)} & \small{(0.000)} & \small{(0.001)} & & \\
\bottomrule
\end{tabular*}
\begin{tablenotes}[para,flushleft]
\footnotesize{\textit{Notes}: Bold face indicates the best classifier (i.e., with the lowest average rank over the data sets) per performance metric. The last row shows the Friedman statistic with p-values. Underlined values indicate that a classifier performs significantly worse than the best classifier in a data scenario at the 5 percent level according to the Nemenyi test or, in the case of the Friedman statistic, indicate that not all classifiers perform equally well.}
\end{tablenotes}
\end{threeparttable}
\end{center}
\label{table:results6}
\end{table*}

However, comparing the different sampling strategies yields some surprising results. 
Table \ref{table:results7} displays the aggregated, averaged ranks of the classifiers in the six different data scenarios. 
The second to last line, AvgR, summarizes the average rank of the sampling method, which is reflected in the last line reporting the overall rank of a sampling strategy.
Remarkably, up-sampling and down-sampling, the two simplest sampling strategies, outperform the more sophisticated competing sampling methods. 
BSMOTE ranks third, ROSE fourth, and SMOTE is even outranked by the original data scenario, i.e., no sampling method at all. 
This implies that the development and usage of more complicated sampling methods may not be worthwhile. 
It also contradicts the finding of \citet{Drummond2003}, who compare the performance of C4.5 decision trees on under-sampled and up-sampled data and conclude that under-sampling is preferable to up-sampling. 
However, they only use one performance measure and one classifier.
Thus, the results of the present study might be more robust.

\begin{table*}[t]
\caption{Results sampling methods comparison} \vspace*{-1.5em}\small
\begin{center}
\begin{threeparttable}
\begin{tabular*}{\textwidth}{@{\extracolsep{\fill}} lcccccc}
\toprule
 Data set & original & down-sampling & up-sampling & SMOTE & ROSE & BSMOTE\\ 
  \midrule
LDA & 4.19 & 3.42 & \textbf{2.08} & 4.40 & 3.04 & 3.88 \\ 
  QDA & 4.25 & 2.75 & \textbf{2.46} & 4.96 & 2.71 & 3.88 \\ 
  FDA & 3.08 & 2.79 & \textbf{2.46} & 3.67 & 4.92 & 4.08 \\ 
  LogReg & 4.92 & 2.23 & \textbf{2.10} & 4.88 & 3.12 & 3.75 \\ 
  SVM-L & 4.65 & 3.27 & \textbf{1.65} & 4.88 & 2.94 & 3.62 \\ 
  SVM-R & 4.56 & \textbf{1.52} & 1.94 & 5.50 & 3.19 & 4.29 \\ 
  KNN & 3.92 & \textbf{2.42} & 2.96 & 4.67 & 3.75 & 3.29 \\ 
  ANN & 3.71 & \textbf{2.46} & 3.08 & 4.92 & 3.21 & 3.62 \\ 
  CART & 3.62 & 3.27 & \textbf{2.69} & 3.85 & 4.50 & 3.06 \\ 
  NB & 4.73 & 3.12 & 2.73 & 5.08 & \textbf{1.46} & 3.88 \\ 
  \midrule
  Bag-CT & 3.94 & 2.83 & \textbf{2.58} & 4.23 & 3.79 & 3.62 \\ 
  Boost-CT & 3.48 & \textbf{2.27} & 2.62 & 3.96 & 4.67 & 4.00 \\ 
  AdaBoost & 3.67 & \textbf{1.75} & 2.38 & 4.33 & 4.54 & 4.33 \\ 
  SGB & 3.83 & 2.29 & \textbf{1.94} & 4.54 & 4.12 & 4.27 \\ 
  RF & 4.33 & 2.54 & \textbf{2.21} & 4.62 & 4.54 & 2.75 \\ 
  parRF & 4.25 & 2.31 & \textbf{2.02} & 4.75 & 4.58 & 3.08 \\ 
  rotForest & 4.42 & 3.21 &\textbf{1.79} & 4.12 & 4.08 & 3.38 \\ 
  avNNet & 4.54 & \textbf{2.10} & 2.65 & 4.62 & 3.33 & 3.75 \\ 
  Logit-Boost & 4.12 & 3.62 & \textbf{2.25} & 3.71 & 4.00 & 3.29 \\ 
  bagFDA & 4.25 & 2.15 & \textbf{1.85} & 4.96 & 4.21 & 3.58 \\ 
  \midrule
  AvgS & 4.08 & 2.42 & \textbf{2.25} & 4.50 & 4.17 & 3.58 \\ 
  AvgW & 4.06 & 2.46 & \textbf{2.25} & 4.56 & 4.25 & 3.42 \\ 
  Stack & 3.85 & \textbf{2.29} & 2.56 & 5.04 & 4.17 & 3.08 \\
  \midrule
  AvgR & 4.11 & 2.59 & \textbf{2.33} & 4.55 & 3.80 & 3.63 \\ 
  Rank & 5.00 & 2.00 & 1.00 & 6.00 & 4.00 & 3.00 \\ 
\bottomrule
\end{tabular*}
\begin{tablenotes}[para,flushleft]
\footnotesize{\textit{Notes}: Bold face indicates the best performance of a classifier (i.e., lowest average rank over all data sets and all performance metrics) with respect to different sampling strategies.}
\end{tablenotes}
\end{threeparttable}
\end{center}
\label{table:results7}
\end{table*}

Concerning the classifiers' relative performance, the results per data scenario show that the relative rank of a model hardly changes with differently sampled data, especially amongst the best models. Therefore, a model that performs well on balanced data, like RF or a boosting model, also performs relatively well on unbalanced data. Overall, it can be concluded that choosing RF or a boosting method in combination with up-sampling seems to be a good strategy in a credit scoring application with unequal class distributions.

\section{Conclusion}
This study evaluates 23 classification strategies in the credit scoring context on four different data sets in combination with five different sampling strategies addressing class imbalances by utilizing six different performance measures.
The predictive exercise shows that ensemble methods are clearly superior to most individual models. 
Especially boosting or random forest methods perform well throughout all data sampling scenarios. 
While only three heterogeneous ensembles are included in this study, results by \citet{Lessmann2015} indicate that sophisticated heterogeneous models may still outperform the best performing homogeneous ensembles.
However, homogeneous ensemble approaches such as AdaBoost or random forests have the advantage that they not only perform relatively well, but are also readily implemented within different computational platforms and are very well-documented and researched methods. 
More complicated heterogeneous models are still much harder to implement and tune properly. 
However, they provide a fruitful avenue for further research.

Moreover, this paper shows that well-performing classifiers should be combined with simple up-sampling or down-sampling in order to obtain the best possible predictive performance in the face of unequal class distributions. 
This result is somewhat surprising, considering the diversity of other, more sophisticated sampling methods. 
It highlights, however, the importance of using a variety of models and performance measures when evaluating a sampling strategy on a specific classification problem. 
Using only a limited number of classification strategies and performance measures may not give a robust enough picture of a sampling strategy's impact.
Overall, the results of this study support other findings in the literature that there are several easily accessible, relatively simple to implement alternatives to current industry standards like logistic regression or discriminant analysis, which would improve classifications in the credit scoring application, especially in combination with suitable data sampling strategies. 
With increasing interest in machine learning and data mining methods, further advances can be expected and may prove to be a very fruitful research area for both academics and professionals concerned with credit scoring.
For further research, the inclusion of more heterogeneous ensembles in combination with data sampling may be a promising way to find optimal strategies in the credit scoring context. 
It may also be worth to further explore differences in the models' implementations, not only in terms of accessibility and handling, but also in terms of performance, as \citet{Fernandez-Delgado:2014} found that performance differences may not only exist between models, but also between one model's different implementations.

\small{\scfont\setstretch{0.85}
\addcontentsline{toc}{section}{References}
\bibliographystyle{custom.bst}
\bibliography{lit}}\normalsize

\clearpage
\begin{appendices}\crefalias{section}{appsec}
\setcounter{equation}{0}
\renewcommand\theequation{A.\arabic{equation}}
\section{Additional results}\label{app:additionalres}

\begin{table}[h] \centering 
  \caption{GMSC summary statistics of original data sets and subsets} 
  \label{tab_app:data1}
\scalebox{0.8}{
\begin{tabular}{@{\extracolsep{5pt}}llcccccc} 
\\[-1.8ex]\hline 
\hline \\[-1.8ex] 
Statistic & & \multicolumn{1}{c}{Mean} & \multicolumn{1}{c}{St. Dev.} & \multicolumn{1}{c}{Min} & \multicolumn{1}{c}{Pctl(25)} & \multicolumn{1}{c}{Pctl(75)} & \multicolumn{1}{c}{Max} \\ 
\hline \\[-1.8ex] 
\multirow{10}{*}{GMSC} & RevolvingUtilizationOfUnsecuredLines & 6.05 & 249.76 & 0 & 0.03 & 0.56 & 50,708.00 \\ 
& age & 52.30 & 14.77 & 0 & 41 & 63 & 109 \\ 
& No.Time30.59DaysPastDueNotWorse  & 0.42 & 4.19 & 0 & 0 & 0 & 98 \\ 
& DebtRatio & 353.01 & 2,037.82 & 0 & 0.18 & 0.87 & 329,664.00 \\ 
& MonthlyIncome  & 6,670.22 & 14,384.67 & 0 & 3,400.00 & 8,249.00 & 3,008,750.00 \\ 
& No.OpenCreditLinesAndLoans  & 8.45 & 5.15 & 0 & 5 & 11 & 58 \\ 
& No.Times90DaysLate  & 0.27 & 4.17 & 0 & 0 & 0 & 98 \\ 
& NumberRealEstateLoansOrLines  & 1.02 & 1.13 & 0 & 0 & 2 & 54 \\ 
& No.Time60.89DaysPastDueNotWorse  & 0.24 & 4.16 & 0 & 0 & 0 & 98 \\ 
& No.Dependents & 0.76 & 1.12 & 0 & 0 & 1.00 & 20.00 \\ 
\hline \\[-1.8ex] 
\multirow{10}{*}{\shortstack{GMSC \\ small}} & RevolvingUtilizationOfUnsecuredLines & 5.18 & 150.39 & 0 & 0.03 & 0.54 & 8,497.00 \\ 
& age & 52.43 & 14.91 & 0 & 41 & 63 & 101 \\ 
& No.Time30.59DaysPastDueNotWorse & 0.35 & 3.26 & 0 & 0 & 0 & 98 \\ 
& DebtRatio  & 355.05 & 1,409.41 & 0 & 0.18 & 0.88 & 61,907.00 \\ 
& MonthlyIncome & 6,472.04 & 5,617.28 & 0 & 3,333.00 & 8,244.00 & 184,903.00 \\ 
& No.OpenCreditLinesAndLoans  & 8.51 & 5.12 & 0 & 5 & 11 & 48 \\ 
& No.Times90DaysLate  & 0.19 & 3.23 & 0 & 0 & 0 & 98 \\ 
& NumberRealEstateLoansOrLines & 1.03 & 1.15 & 0 & 0 & 2 & 23 \\ 
& No.Time60.89DaysPastDueNotWorse  & 0.16 & 3.21 & 0 & 0 & 0 & 98 \\ 
& No.Dependents  & 0.75 & 1.10 & 0 & 0 & 1.00 & 8.00 \\ 
\\[-1.8ex]\hline 
\hline \\[-1.8ex]
\end{tabular}}
\end{table} 

\begin{table}[t] \centering 
  \caption{TC summary statistics of original data and subset} 
  \label{tab_app:data2}
\scalebox{0.8}{
\begin{tabular}{@{\extracolsep{5pt}}llcccccc} 
\\[-1.8ex]\hline 
\hline \\[-1.8ex] 
Statistic & & \multicolumn{1}{c}{Mean} & \multicolumn{1}{c}{St. Dev.} & \multicolumn{1}{c}{Min} & \multicolumn{1}{c}{Pctl(25)} & \multicolumn{1}{c}{Pctl(75)} & \multicolumn{1}{c}{Max} \\ 
\hline \\[-1.8ex] 
\multirow{25}{*}{TC}& LIMIT\_BAL  & 167,484.30 & 129,747.70 & 10,000 & 50,000 & 240,000 & 1,000,000 \\ 
& SEX & 1.60 & 0.49 & 1 & 1 & 2 & 2 \\ 
& EDUCATION  & 1.85 & 0.79 & 0 & 1 & 2 & 6 \\ 
& MARRIAGE  & 1.55 & 0.52 & 0 & 1 & 2 & 3 \\ 
& AGE  & 35.49 & 9.22 & 21 & 28 & 41 & 79 \\ 
& PAY\_0 & $-$0.02 & 1.12 & $-$2 & $-$1 & 0 & 8 \\ 
& PAY\_2  & $-$0.13 & 1.20 & $-$2 & $-$1 & 0 & 8 \\ 
& PAY\_3  & $-$0.17 & 1.20 & $-$2 & $-$1 & 0 & 8 \\ 
& PAY\_4 & $-$0.22 & 1.17 & $-$2 & $-$1 & 0 & 8 \\ 
& PAY\_5  & $-$0.27 & 1.13 & $-$2 & $-$1 & 0 & 8 \\ 
& PAY\_6  & $-$0.29 & 1.15 & $-$2 & $-$1 & 0 & 8 \\ 
& BILL\_AMT1  & 51,223.33 & 73,635.86 & $-$165,580 & 3,558.8 & 67,091 & 964,511 \\ 
& BILL\_AMT2  & 49,179.08 & 71,173.77 & $-$69,777 & 2,984.8 & 64,006.2 & 983,931 \\ 
& BILL\_AMT3  & 47,013.15 & 69,349.39 & $-$157,264 & 2,666.2 & 60,164.8 & 1,664,089 \\ 
& BILL\_AMT4 & 43,262.95 & 64,332.86 & $-$170,000 & 2,326.8 & 54,506 & 891,586 \\ 
& BILL\_AMT5  & 40,311.40 & 60,797.16 & $-$81,334 & 1,763 & 50,190.5 & 927,171 \\ 
& BILL\_AMT6  & 38,871.76 & 59,554.11 & $-$339,603 & 1,256 & 49,198.2 & 961,664 \\ 
& PAY\_AMT1  & 5,663.58 & 16,563.28 & 0 & 1,000 & 5,006 & 873,552 \\ 
& PAY\_AMT2 & 5,921.16 & 23,040.87 & 0 & 833 & 5,000 & 1,684,259 \\ 
& PAY\_AMT3  & 5,225.68 & 17,606.96 & 0 & 390 & 4,505 & 896,040 \\ 
& PAY\_AMT4  & 4,826.08 & 15,666.16 & 0 & 296 & 4,013.2 & 621,000 \\ 
& PAY\_AMT5  & 4,799.39 & 15,278.31 & 0 & 252.5 & 4,031.5 & 426,529 \\ 
& PAY\_AMT6  & 5,215.50 & 17,777.47 & 0 & 117.8 & 4,000 & 528,666 \\
\hline \\[-1.8ex] 
\multirow{25}{*}{\shortstack{TC \\ small}} & LIMIT\_BAL & 172,585.80 & 132,972.60 & 10,000 & 60,000 & 240,000 & 800,000 \\ 
& SEX & 1.61 & 0.49 & 1 & 1 & 2 & 2 \\ 
& EDUCATION & 1.85 & 0.77 & 0 & 1 & 2 & 6 \\ 
& MARRIAGE & 1.55 & 0.52 & 0 & 1 & 2 & 3 \\ 
& AGE & 35.30 & 9.15 & 21 & 28 & 41 & 75 \\ 
& PAY\_0  & $-$0.03 & 1.10 & $-$2 & $-$1 & 0 & 8 \\ 
& PAY\_2  & $-$0.14 & 1.16 & $-$2 & $-$1 & 0 & 7 \\ 
& PAY\_3  & $-$0.18 & 1.15 & $-$2 & $-$1 & 0 & 7 \\ 
& PAY\_4  & $-$0.22 & 1.13 & $-$2 & $-$1 & 0 & 7 \\ 
& PAY\_5  & $-$0.27 & 1.10 & $-$2 & $-$1 & 0 & 7 \\ 
& PAY\_6  & $-$0.30 & 1.11 & $-$2 & $-$1 & 0 & 7 \\ 
& BILL\_AMT1  & 52,683.07 & 75,001.30 & $-$15,308 & 3,814 & 70,327 & 608,594 \\ 
& BILL\_AMT2  & 50,146.48 & 72,534.64 & $-$69,777 & 3,362 & 66,087 & 624,475 \\ 
& BILL\_AMT3  & 48,803.24 & 72,489.07 & $-$25,443 & 2,987 & 61,465 & 689,643 \\ 
& BILL\_AMT4  & 44,550.35 & 66,037.63 & $-$46,627 & 2,400 & 57,564 & 616,836 \\ 
& BILL\_AMT5  & 41,344.53 & 63,232.66 & $-$46,627 & 1,771 & 52,175 & 823,540 \\ 
& BILL\_AMT6  & 39,824.10 & 61,397.40 & $-$46,627 & 1,200 & 49,880 & 501,370 \\ 
& PAY\_AMT1  & 5,725.00 & 17,043.14 & 0 & 1,000 & 5,025 & 368,199 \\ 
& PAY\_AMT2 & 6,424.60 & 21,659.56 & 0 & 1,000 & 5,000 & 401,003 \\ 
& PAY\_AMT3  & 5,011.26 & 14,477.74 & 0 & 498 & 4,394 & 245,863 \\ 
& PAY\_AMT4  & 4,764.35 & 14,466.55 & 0 & 286 & 4,000 & 232,242 \\ 
& PAY\_AMT5  & 5,129.95 & 18,604.51 & 0 & 249 & 4,000 & 417,990 \\ 
& PAY\_AMT6 1 & 5,308.94 & 17,951.69 & 0 & 184 & 4,019 & 372,495 \\ 
\\[-1.8ex]\hline 
\hline \\[-1.8ex]
\end{tabular}}
\end{table} 

\begin{sidewaystable}
\centering
\caption{Results for Accuracy per model and data set}
\label{results:details_accuracy}
\scalebox{0.75}{
\begin{tabular}{lrrrr|rrrr|rrrr|rrrr|rrrr|rrrr}
 \\[-1.8ex]\hline 
\hline \\[-1.8ex]
  Data set & \multicolumn{4}{c|}{original} & \multicolumn{4}{c|}{down-sampling} & \multicolumn{4}{c|}{up-sampling} & \multicolumn{4}{c|}{SMOTE} & \multicolumn{4}{c|}{ROSE} & \multicolumn{4}{c}{BSMOTE}\\
 & GC & AC & TC & GMSC & GC & AC & TC & GMSC & GC & AC & TC & GMSC & GC & AC & TC & GMSC & GC & AC & TC & GMSC & GC & AC & TC & GMSC \\ 
  \hline
LDA & 0.57 & 0.82 & 0.37 & 0.13 & 0.68 & 0.85 & 0.57 & 0.53 & 0.85 & 0.85 & 0.57 & 0.53 & 0.62 & 0.80 & 0.51 & 0.13 & 0.66 & 0.84 & 0.57 & 0.53 & 0.66 & 0.85 & 0.54 & 0.53 \\ 
  QDA & 0.54 & 0.82 & 0.39 & 0.13 & 0.64 & 0.84 & 0.59 & 0.55 & 0.85 & 0.85 & 0.59 & 0.54 & 0.60 & 0.81 & 0.54 & 0.13 & 0.64 & 0.84 & 0.58 & 0.53 & 0.64 & 0.85 & 0.56 & 0.53 \\ 
  FDA & 0.56 & 0.85 & 0.68 & 0.61 & 0.66 & 0.85 & 0.61 & 0.55 & 0.86 & 0.86 & 0.60 & 0.54 & 0.64 & 0.85 & 0.56 & 0.61 & 0.62 & 0.84 & 0.59 & 0.51 & 0.62 & 0.84 & 0.58 & 0.54 \\ 
  LogReg & 0.57 & 0.82 & 0.37 & 0.13 & 0.68 & 0.85 & 0.57 & 0.54 & 0.85 & 0.85 & 0.57 & 0.54 & 0.64 & 0.79 & 0.51 & 0.13 & 0.67 & 0.84 & 0.57 & 0.53 & 0.66 & 0.84 & 0.54 & 0.54 \\ 
  SVM-L & 0.57 & 0.82 & 0.37 & 0.13 & 0.68 & 0.85 & 0.56 & 0.54 & 0.85 & 0.85 & 0.58 & 0.54 & 0.62 & 0.80 & 0.51 & 0.13 & 0.68 & 0.84 & 0.56 & 0.53 & 0.66 & 0.85 & 0.55 & 0.54 \\ 
  SVM-R & 0.54 & 0.82 & 0.38 & 0.13 & 0.69 & 0.86 & 0.60 & 0.55 & 0.86 & 0.86 & 0.60 & 0.54 & 0.58 & 0.79 & 0.52 & 0.13 & 0.64 & 0.84 & 0.59 & 0.53 & 0.61 & 0.84 & 0.56 & 0.53 \\ 
  KNN & 0.53 & 0.78 & 0.51 & 0.65 & 0.62 & 0.87 & 0.64 & 0.53 & 0.87 & 0.87 & 0.70 & 0.90 & 0.61 & 0.44 & 0.63 & 0.65 & 0.60 & 0.84 & 0.68 & 0.52 & 0.62 & 0.87 & 0.55 & 0.79 \\ 
  ANN & 0.56 & 0.83 & 0.41 & 0.13 & 0.67 & 0.85 & 0.60 & 0.55 & 0.79 & 0.79 & 0.61 & 0.54 & 0.62 & 0.81 & 0.51 & 0.13 & 0.66 & 0.85 & 0.60 & 0.53 & 0.63 & 0.85 & 0.55 & 0.54 \\ 
  CART & 0.67 & 0.44 & 0.79 & 0.91 & 0.68 & 0.85 & 0.71 & 0.61 & 0.85 & 0.85 & 0.67 & 0.63 & 0.65 & 0.44 & 0.72 & 0.91 & 0.68 & 0.84 & 0.77 & 0.93 & 0.65 & 0.85 & 0.73 & 0.79 \\ 
  NB & 0.56 & 0.76 & 0.39 & 0.13 & 0.68 & 0.80 & 0.59 & 0.54 & 0.79 & 0.79 & 0.59 & 0.53 & 0.64 & 0.74 & 0.53 & 0.13 & 0.68 & 0.82 & 0.60 & 0.54 & 0.65 & 0.80 & 0.56 & 0.52 \\ 
  \hline
  Bag-CT & 0.52 & 0.80 & 0.41 & 0.61 & 0.63 & 0.87 & 0.63 & 0.57 & 0.85 & 0.85 & 0.62 & 0.61 & 0.62 & 0.80 & 0.55 & 0.61 & 0.66 & 0.84 & 0.64 & 0.93 & 0.64 & 0.87 & 0.58 & 0.61 \\ 
  Boost-CT & 0.56 & 0.85 & 0.42 & 0.18 & 0.69 & 0.86 & 0.61 & 0.55 & 0.86 & 0.86 & 0.71 & 0.54 & 0.65 & 0.79 & 0.59 & 0.18 & 0.60 & 0.84 & 0.76 & 0.53 & 0.66 & 0.85 & 0.57 & 0.54 \\ 
  AdaBoost & 0.56 & 0.82 & 0.43 & 0.13 & 0.68 & 0.86 & 0.63 & 0.55 & 0.84 & 0.84 & 0.61 & 0.54 & 0.62 & 0.82 & 0.54 & 0.13 & 0.64 & 0.82 & 0.58 & 0.53 & 0.68 & 0.80 & 0.56 & 0.54 \\ 
  SGB & 0.55 & 0.83 & 0.42 & 0.33 & 0.67 & 0.85 & 0.62 & 0.55 & 0.87 & 0.87 & 0.61 & 0.55 & 0.65 & 0.82 & 0.55 & 0.33 & 0.67 & 0.84 & 0.59 & 0.54 & 0.66 & 0.85 & 0.57 & 0.54 \\ 
  RF & 0.56 & 0.84 & 0.41 & 0.16 & 0.69 & 0.86 & 0.61 & 0.55 & 0.86 & 0.86 & 0.61 & 0.56 & 0.64 & 0.82 & 0.57 & 0.16 & 0.68 & 0.84 & 0.61 & 0.54 & 0.67 & 0.87 & 0.60 & 0.56 \\ 
  parRF & 0.55 & 0.84 & 0.42 & 0.21 & 0.69 & 0.87 & 0.61 & 0.55 & 0.87 & 0.87 & 0.61 & 0.56 & 0.64 & 0.82 & 0.57 & 0.21 & 0.68 & 0.84 & 0.61 & 0.54 & 0.65 & 0.85 & 0.60 & 0.56 \\ 
  rotRF & 0.54 & 0.44 & 0.78 & 0.93 & 0.70 & 0.85 & 0.76 & 0.55 & 0.86 & 0.86 & 0.60 & 0.68 & 0.63 & 0.84 & 0.56 & 0.93 & 0.67 & 0.84 & 0.59 & 0.58 & 0.63 & 0.87 & 0.59 & 0.55 \\ 
  avNNet & 0.55 & 0.82 & 0.40 & 0.13 & 0.68 & 0.85 & 0.61 & 0.54 & 0.85 & 0.85 & 0.60 & 0.54 & 0.64 & 0.81 & 0.54 & 0.13 & 0.68 & 0.84 & 0.60 & 0.54 & 0.64 & 0.87 & 0.57 & 0.54 \\ 
  Logit-Boost & 0.57 & 0.75 & 0.68 & 0.81 & 0.62 & 0.85 & 0.71 & 0.77 & 0.78 & 0.78 & 0.75 & 0.78 & 0.62 & 0.76 & 0.77 & 0.81 & 0.66 & 0.78 & 0.72 & 0.76 & 0.72 & 0.82 & 0.75 & 0.76 \\ 
  bagFDA & 0.56 & 0.84 & 0.42 & 0.13 & 0.70 & 0.85 & 0.62 & 0.55 & 0.85 & 0.85 & 0.61 & 0.54 & 0.65 & 0.82 & 0.54 & 0.13 & 0.66 & 0.85 & 0.59 & 0.52 & 0.67 & 0.84 & 0.57 & 0.54 \\ 
  \hline
  AvgS & 0.55 & 0.82 & 0.41 & 0.13 & 0.68 & 0.85 & 0.62 & 0.55 & 0.86 & 0.86 & 0.61 & 0.55 & 0.64 & 0.80 & 0.55 & 0.13 & 0.66 & 0.84 & 0.61 & 0.53 & 0.67 & 0.87 & 0.60 & 0.54 \\ 
  AvgW & 0.55 & 0.82 & 0.41 & 0.13 & 0.68 & 0.85 & 0.62 & 0.55 & 0.86 & 0.86 & 0.61 & 0.55 & 0.64 & 0.80 & 0.55 & 0.13 & 0.66 & 0.84 & 0.61 & 0.53 & 0.67 & 0.87 & 0.60 & 0.54 \\ 
  Stack & 0.56 & 0.82 & 0.41 & 0.37 & 0.67 & 0.86 & 0.62 & 0.55 & 0.85 & 0.85 & 0.61 & 0.87 & 0.63 & 0.79 & 0.54 & 0.37 & 0.64 & 0.83 & 0.60 & 0.54 & 0.66 & 0.87 & 0.58 & 0.55 \\ 
 \\[-1.8ex]\hline 
\hline \\[-1.8ex]
\end{tabular}}
\end{sidewaystable}

\begin{sidewaystable}
    \centering
\caption{Results for $\kappa$ per model and data set}
\label{results:details_kappa}
\scalebox{0.75}{
\begin{tabular}{lrrrr|rrrr|rrrr|rrrr|rrrr|rrrr}
 \\[-1.8ex]\hline 
\hline \\[-1.8ex]
  Data set & \multicolumn{4}{c|}{original} & \multicolumn{4}{c|}{down-sampling} & \multicolumn{4}{c|}{up-sampling} & \multicolumn{4}{c|}{SMOTE} & \multicolumn{4}{c|}{ROSE} & \multicolumn{4}{c}{BSMOTE}\\
 & GC & AC & TC & GMSC & GC & AC & TC & GMSC & GC & AC & TC & GMSC & GC & AC & TC & GMSC & GC & AC & TC & GMSC & GC & AC & TC & GMSC \\ 
  \hline
LDA & 0.26 & 0.64 & 0.04 & 0.01 & 0.37 & 0.70 & 0.14 & 0.06 & 0.70 & 0.70 & 0.13 & 0.06 & 0.28 & 0.61 & 0.10 & 0.01 & 0.32 & 0.68 & 0.13 & 0.06 & 0.33 & 0.71 & 0.12 & 0.05 \\ 
  QDA & 0.20 & 0.64 & 0.08 & 0.00 & 0.29 & 0.67 & 0.19 & 0.09 & 0.70 & 0.70 & 0.19 & 0.07 & 0.25 & 0.63 & 0.14 & 0.00 & 0.27 & 0.68 & 0.17 & 0.07 & 0.29 & 0.71 & 0.15 & 0.06 \\ 
  FDA & 0.23 & 0.71 & 0.24 & 0.09 & 0.30 & 0.71 & 0.21 & 0.09 & 0.72 & 0.72 & 0.19 & 0.08 & 0.32 & 0.70 & 0.18 & 0.09 & 0.24 & 0.68 & 0.18 & 0.02 & 0.26 & 0.68 & 0.19 & 0.08 \\ 
  LogReg & 0.26 & 0.64 & 0.04 & 0.01 & 0.37 & 0.70 & 0.14 & 0.08 & 0.70 & 0.70 & 0.13 & 0.07 & 0.31 & 0.59 & 0.10 & 0.01 & 0.34 & 0.68 & 0.13 & 0.07 & 0.33 & 0.68 & 0.12 & 0.08 \\ 
  SVM-L & 0.26 & 0.64 & 0.03 & 0.00 & 0.35 & 0.70 & 0.12 & 0.07 & 0.70 & 0.70 & 0.16 & 0.08 & 0.28 & 0.61 & 0.10 & 0.00 & 0.37 & 0.68 & 0.13 & 0.07 & 0.33 & 0.71 & 0.12 & 0.08 \\ 
  SVM-R & 0.20 & 0.64 & 0.05 & 0.00 & 0.38 & 0.72 & 0.20 & 0.10 & 0.72 & 0.72 & 0.20 & 0.08 & 0.20 & 0.59 & 0.12 & 0.00 & 0.29 & 0.68 & 0.18 & 0.06 & 0.24 & 0.68 & 0.16 & 0.06 \\ 
  KNN & 0.18 & 0.58 & 0.14 & 0.05 & 0.24 & 0.73 & 0.23 & 0.05 & 0.73 & 0.73 & 0.12 & 0.05 & 0.15 & 0.00 & 0.13 & 0.05 & 0.12 & 0.68 & 0.16 & 0.03 & 0.13 & 0.73 & 0.13 & 0.08 \\ 
  ANN & 0.23 & 0.66 & 0.09 & 0.01 & 0.33 & 0.70 & 0.20 & 0.10 & 0.58 & 0.58 & 0.21 & 0.09 & 0.28 & 0.63 & 0.10 & 0.01 & 0.32 & 0.70 & 0.20 & 0.07 & 0.27 & 0.71 & 0.14 & 0.09 \\ 
  CART & 0.34 & 0.00 & 0.31 & 0.22 & 0.34 & 0.71 & 0.28 & 0.12 & 0.71 & 0.71 & 0.26 & 0.13 & 0.31 & -0.00 & 0.29 & 0.22 & 0.29 & 0.68 & 0.33 & 0.00 & 0.30 & 0.71 & 0.31 & 0.19 \\ 
  NB & 0.25 & 0.53 & 0.07 & 0.00 & 0.35 & 0.60 & 0.17 & 0.07 & 0.58 & 0.58 & 0.18 & 0.05 & 0.32 & 0.49 & 0.13 & 0.00 & 0.35 & 0.65 & 0.20 & 0.08 & 0.32 & 0.59 & 0.15 & 0.05 \\ 
  \hline
  Bag-CT & 0.18 & 0.61 & 0.10 & 0.11 & 0.25 & 0.74 & 0.23 & 0.10 & 0.71 & 0.71 & 0.20 & 0.10 & 0.28 & 0.61 & 0.14 & 0.11 & 0.32 & 0.68 & 0.23 & 0.00 & 0.28 & 0.73 & 0.17 & 0.12 \\ 
  Boost-CT & 0.23 & 0.71 & 0.11 & 0.02 & 0.38 & 0.72 & 0.23 & 0.10 & 0.72 & 0.72 & 0.27 & 0.08 & 0.34 & 0.59 & 0.20 & 0.02 & 0.21 & 0.68 & 0.32 & 0.06 & 0.33 & 0.71 & 0.18 & 0.08 \\ 
  AdaBoost & 0.25 & 0.64 & 0.12 & 0.01 & 0.35 & 0.72 & 0.25 & 0.10 & 0.67 & 0.67 & 0.21 & 0.09 & 0.28 & 0.65 & 0.15 & 0.01 & 0.27 & 0.65 & 0.17 & 0.05 & 0.38 & 0.59 & 0.16 & 0.09 \\ 
  SGB & 0.22 & 0.66 & 0.12 & 0.04 & 0.33 & 0.70 & 0.23 & 0.10 & 0.74 & 0.74 & 0.22 & 0.09 & 0.34 & 0.65 & 0.16 & 0.04 & 0.34 & 0.68 & 0.18 & 0.09 & 0.33 & 0.71 & 0.17 & 0.09 \\ 
  RF & 0.23 & 0.69 & 0.10 & 0.01 & 0.38 & 0.72 & 0.21 & 0.10 & 0.72 & 0.72 & 0.21 & 0.10 & 0.32 & 0.65 & 0.19 & 0.01 & 0.37 & 0.68 & 0.21 & 0.05 & 0.36 & 0.73 & 0.22 & 0.11 \\ 
  parRF & 0.22 & 0.69 & 0.11 & 0.02 & 0.38 & 0.74 & 0.21 & 0.10 & 0.74 & 0.74 & 0.22 & 0.10 & 0.32 & 0.65 & 0.19 & 0.02 & 0.35 & 0.68 & 0.21 & 0.04 & 0.32 & 0.71 & 0.22 & 0.11 \\ 
  rotRF & 0.20 & 0.00 & 0.32 & 0.00 & 0.40 & 0.71 & 0.32 & 0.09 & 0.72 & 0.72 & 0.19 & 0.14 & 0.29 & 0.68 & 0.19 & 0.00 & 0.34 & 0.68 & 0.18 & 0.05 & 0.27 & 0.73 & 0.20 & 0.10 \\ 
  avNNet & 0.22 & 0.64 & 0.08 & 0.01 & 0.37 & 0.70 & 0.22 & 0.08 & 0.70 & 0.70 & 0.20 & 0.08 & 0.31 & 0.63 & 0.15 & 0.01 & 0.35 & 0.68 & 0.20 & 0.08 & 0.30 & 0.73 & 0.18 & 0.08 \\ 
  Logit-Boost & 0.23 & 0.52 & 0.25 & 0.19 & 0.21 & 0.70 & 0.25 & 0.18 & 0.58 & 0.58 & 0.31 & 0.17 & 0.27 & 0.53 & 0.27 & 0.19 & 0.27 & 0.57 & 0.29 & 0.11 & 0.30 & 0.65 & 0.23 & 0.19 \\ 
  bagFDA & 0.25 & 0.69 & 0.11 & 0.01 & 0.40 & 0.70 & 0.25 & 0.10 & 0.70 & 0.70 & 0.21 & 0.09 & 0.34 & 0.65 & 0.15 & 0.01 & 0.32 & 0.70 & 0.18 & 0.04 & 0.35 & 0.68 & 0.17 & 0.09 \\ 
  \hline
  AvgS & 0.22 & 0.64 & 0.10 & 0.01 & 0.35 & 0.70 & 0.24 & 0.10 & 0.72 & 0.72 & 0.23 & 0.10 & 0.31 & 0.61 & 0.17 & 0.01 & 0.32 & 0.68 & 0.22 & 0.07 & 0.35 & 0.73 & 0.22 & 0.08 \\ 
  AvgW & 0.22 & 0.64 & 0.10 & 0.01 & 0.35 & 0.70 & 0.24 & 0.10 & 0.72 & 0.72 & 0.23 & 0.10 & 0.31 & 0.61 & 0.17 & 0.01 & 0.32 & 0.68 & 0.22 & 0.07 & 0.35 & 0.73 & 0.22 & 0.08 \\ 
  Stack & 0.23 & 0.64 & 0.09 & 0.05 & 0.33 & 0.72 & 0.24 & 0.09 & 0.70 & 0.70 & 0.21 & 0.14 & 0.29 & 0.59 & 0.14 & 0.05 & 0.29 & 0.66 & 0.21 & 0.08 & 0.33 & 0.73 & 0.18 & 0.10 \\ 
   \\[-1.8ex]\hline 
\hline \\[-1.8ex]
\end{tabular}}
\end{sidewaystable}

\begin{sidewaystable}
\caption{Results for BS per model and data set}
\label{results:details_BS}
\scalebox{0.75}{
\begin{tabular}{lrrrr|rrrr|rrrr|rrrr|rrrr|rrrr}
 \\[-1.8ex]\hline 
\hline \\[-1.8ex]
  Data set & \multicolumn{4}{c|}{original} & \multicolumn{4}{c|}{down-sampling} & \multicolumn{4}{c|}{up-sampling} & \multicolumn{4}{c|}{SMOTE} & \multicolumn{4}{c|}{ROSE} & \multicolumn{4}{c}{BSMOTE}\\
 & GC & AC & TC & GMSC & GC & AC & TC & GMSC & GC & AC & TC & GMSC & GC & AC & TC & GMSC & GC & AC & TC & GMSC & GC & AC & TC & GMSC \\ 
  \hline
LDA & 0.43 & 0.18 & 0.63 & 0.87 & 0.32 & 0.15 & 0.43 & 0.47 & 0.15 & 0.15 & 0.43 & 0.47 & 0.38 & 0.20 & 0.49 & 0.87 & 0.34 & 0.16 & 0.43 & 0.47 & 0.34 & 0.15 & 0.46 & 0.47 \\ 
  QDA & 0.46 & 0.18 & 0.61 & 0.87 & 0.36 & 0.16 & 0.41 & 0.45 & 0.15 & 0.15 & 0.41 & 0.46 & 0.40 & 0.19 & 0.46 & 0.87 & 0.36 & 0.16 & 0.42 & 0.47 & 0.36 & 0.15 & 0.44 & 0.47 \\ 
  FDA & 0.44 & 0.15 & 0.32 & 0.39 & 0.34 & 0.15 & 0.39 & NA & 0.14 & 0.14 & 0.40 & 0.46 & 0.36 & 0.15 & 0.44 & 0.39 & 0.38 & 0.16 & 0.41 & 0.49 & 0.38 & 0.16 & 0.42 & 0.46 \\ 
  LogReg & 0.43 & 0.18 & 0.63 & 0.87 & 0.32 & 0.15 & 0.43 & 0.46 & 0.15 & 0.15 & 0.43 & 0.46 & 0.36 & 0.21 & 0.49 & 0.87 & 0.33 & 0.16 & 0.43 & 0.47 & 0.34 & 0.16 & 0.46 & 0.46 \\ 
  SVM-L & 0.43 & 0.18 & 0.63 & 0.87 & 0.32 & 0.15 & 0.44 & 0.46 & 0.15 & 0.15 & 0.42 & 0.46 & 0.38 & 0.20 & 0.49 & 0.87 & 0.32 & 0.16 & 0.44 & 0.47 & 0.34 & 0.15 & 0.45 & 0.46 \\ 
  SVM-R & 0.46 & 0.18 & 0.62 & 0.87 & 0.31 & 0.14 & 0.40 & 0.45 & 0.14 & 0.14 & 0.40 & 0.46 & 0.42 & 0.21 & 0.48 & 0.87 & 0.36 & 0.16 & 0.41 & 0.47 & 0.39 & 0.16 & 0.44 & 0.47 \\ 
  KNN & 0.47 & 0.22 & 0.49 & 0.35 & 0.38 & 0.13 & 0.36 & 0.47 & 0.13 & 0.13 & 0.30 & 0.10 & 0.39 & 0.44 & 0.37 & 0.35 & 0.40 & 0.16 & 0.32 & 0.48 & 0.38 & 0.13 & 0.45 & 0.21 \\ 
  ANN & 0.44 & 0.17 & 0.59 & 0.87 & 0.33 & 0.15 & 0.40 & 0.45 & 0.21 & 0.21 & 0.39 & 0.46 & 0.38 & 0.19 & 0.49 & 0.87 & 0.34 & 0.15 & 0.40 & 0.47 & 0.37 & 0.15 & 0.45 & 0.46 \\ 
  CART & 0.33 & 0.44 & 0.21 & 0.09 & 0.32 & 0.15 & 0.29 & 0.39 & 0.15 & 0.15 & 0.33 & 0.37 & 0.35 & 0.56 & 0.28 & 0.09 & 0.32 & 0.16 & 0.23 & 0.93 & 0.35 & 0.15 & 0.27 & 0.21 \\ 
  NB & 0.44 & 0.24 & 0.61 & 0.87 & 0.32 & 0.20 & 0.41 & 0.46 & 0.21 & 0.21 & 0.41 & 0.47 & 0.36 & 0.26 & 0.47 & 0.87 & 0.32 & 0.18 & 0.40 & 0.46 & 0.35 & 0.20 & 0.44 & 0.48 \\ 
  \hline
  Bag-CT & 0.48 & 0.20 & 0.59 & 0.39 & 0.37 & 0.13 & 0.37 & 0.43 & 0.15 & 0.15 & 0.38 & 0.39 & 0.38 & 0.20 & 0.45 & 0.39 & 0.34 & 0.16 & 0.36 & 0.93 & 0.36 & 0.13 & 0.42 & 0.39 \\ 
  Boost-CT & 0.44 & 0.15 & 0.58 & 0.82 & 0.31 & 0.14 & 0.39 & 0.45 & 0.14 & 0.14 & 0.29 & 0.46 & 0.35 & 0.21 & 0.41 & 0.82 & 0.40 & 0.16 & 0.24 & 0.47 & 0.34 & 0.15 & 0.43 & 0.46 \\ 
  AdaBoost & 0.44 & 0.18 & 0.57 & 0.87 & 0.32 & 0.14 & 0.37 & 0.45 & 0.16 & 0.16 & 0.39 & 0.46 & 0.38 & 0.18 & 0.46 & 0.87 & 0.36 & 0.18 & 0.42 & 0.47 & 0.32 & 0.20 & 0.44 & 0.46 \\ 
  SGB & 0.45 & 0.17 & 0.58 & 0.67 & 0.33 & 0.15 & 0.38 & 0.45 & 0.13 & 0.13 & 0.39 & 0.45 & 0.35 & 0.18 & 0.45 & 0.67 & 0.33 & 0.16 & 0.41 & 0.46 & 0.34 & 0.15 & 0.43 & 0.46 \\ 
  RF & 0.44 & 0.16 & 0.59 & 0.84 & 0.31 & 0.14 & 0.39 & 0.45 & 0.14 & 0.14 & 0.39 & 0.44 & 0.36 & 0.18 & 0.43 & 0.84 & 0.32 & 0.16 & 0.39 & 0.46 & 0.33 & 0.13 & 0.40 & 0.44 \\ 
  parRF & 0.45 & 0.16 & 0.58 & 0.79 & 0.31 & 0.13 & 0.39 & 0.45 & 0.13 & 0.13 & 0.39 & 0.44 & 0.36 & 0.18 & 0.43 & 0.79 & 0.32 & 0.16 & 0.39 & 0.46 & 0.35 & 0.15 & 0.40 & 0.44 \\ 
  rotRF & 0.46 & 0.44 & 0.22 & 0.93 & 0.30 & 0.15 & 0.24 & 0.45 & 0.14 & 0.14 & 0.40 & 0.32 & 0.37 & 0.16 & 0.44 & 0.93 & 0.33 & 0.16 & 0.41 & 0.42 & 0.37 & 0.13 & 0.41 & 0.45 \\ 
  avNNet & 0.45 & 0.18 & 0.60 & 0.87 & 0.32 & 0.15 & 0.39 & 0.46 & 0.15 & 0.15 & 0.40 & 0.46 & 0.36 & 0.19 & 0.46 & 0.87 & 0.32 & 0.16 & 0.40 & 0.46 & 0.36 & 0.13 & 0.43 & 0.46 \\ 
  Logit-Boost & 0.43 & 0.25 & 0.32 & 0.19 & 0.38 & 0.15 & 0.29 & 0.23 & 0.22 & 0.22 & 0.25 & 0.22 & 0.38 & 0.24 & 0.23 & 0.19 & 0.34 & 0.22 & 0.28 & 0.24 & 0.28 & 0.18 & 0.25 & 0.24 \\ 
  bagFDA & 0.44 & 0.16 & 0.58 & 0.87 & 0.30 & 0.15 & 0.38 & 0.45 & 0.15 & 0.15 & 0.39 & 0.46 & 0.35 & 0.18 & 0.46 & 0.87 & 0.34 & 0.15 & 0.41 & 0.48 & 0.33 & 0.16 & 0.43 & 0.46 \\ 
  \hline
  AvgS & 0.45 & 0.18 & 0.59 & 0.87 & 0.32 & 0.15 & 0.38 & NA & 0.14 & 0.14 & 0.39 & 0.45 & 0.36 & 0.20 & 0.45 & 0.87 & 0.34 & 0.16 & 0.39 & 0.47 & 0.33 & 0.13 & 0.40 & 0.46 \\ 
  AvgW & 0.45 & 0.18 & 0.59 & 0.87 & 0.32 & 0.15 & 0.38 & NA & 0.14 & 0.14 & 0.39 & 0.45 & 0.36 & 0.20 & 0.45 & 0.87 & 0.34 & 0.16 & 0.39 & 0.47 & 0.33 & 0.13 & 0.40 & 0.46 \\ 
  Stack & 0.44 & 0.18 & 0.59 & 0.63 & 0.33 & 0.14 & 0.38 & NA & 0.15 & 0.15 & 0.39 & 0.13 & 0.37 & 0.21 & 0.46 & 0.63 & 0.36 & 0.17 & 0.40 & 0.46 & 0.34 & 0.13 & 0.42 & 0.45 \\ 
  \\[-1.8ex]\hline 
\hline \\[-1.8ex]
\end{tabular}}
\end{sidewaystable}

\begin{sidewaystable}
\caption{Results for AUC per model and data set}
\label{results:details_AUC}
\scalebox{0.75}{
\begin{tabular}{lrrrr|rrrr|rrrr|rrrr|rrrr|rrrr}
 \\[-1.8ex]\hline 
\hline \\[-1.8ex]
  Data set & \multicolumn{4}{c|}{original} & \multicolumn{4}{c|}{down-sampling} & \multicolumn{4}{c|}{up-sampling} & \multicolumn{4}{c|}{SMOTE} & \multicolumn{4}{c|}{ROSE} & \multicolumn{4}{c}{BSMOTE}\\
 & GC & AC & TC & GMSC & GC & AC & TC & GMSC & GC & AC & TC & GMSC & GC & AC & TC & GMSC & GC & AC & TC & GMSC & GC & AC & TC & GMSC \\ 
  \hline
LDA & 0.81 & 0.91 & 0.66 & 0.68 & 0.81 & 0.91 & 0.67 & 0.66 & 0.92 & 0.92 & 0.67 & 0.66 & 0.80 & 0.91 & 0.67 & 0.68 & 0.80 & 0.92 & 0.68 & 0.65 & 0.80 & 0.92 & 0.67 & 0.64 \\ 
  QDA & 0.76 & 0.89 & 0.69 & 0.70 & 0.76 & 0.89 & 0.68 & 0.77 & 0.89 & 0.89 & 0.68 & 0.70 & 0.75 & 0.89 & 0.68 & 0.70 & 0.75 & 0.89 & 0.69 & 0.71 & 0.74 & 0.89 & 0.68 & 0.67 \\ 
  FDA & 0.78 & 0.95 & 0.68 & 0.71 & 0.73 & 0.93 & 0.74 & 0.77 & 0.95 & 0.95 & 0.71 & 0.76 & 0.75 & 0.94 & 0.71 & 0.71 & 0.76 & 0.91 & 0.71 & 0.54 & 0.75 & 0.91 & 0.72 & 0.76 \\ 
  LogReg & 0.80 & 0.91 & 0.67 & 0.66 & 0.80 & 0.92 & 0.67 & 0.75 & 0.92 & 0.92 & 0.67 & 0.75 & 0.80 & 0.91 & 0.67 & 0.66 & 0.80 & 0.92 & 0.68 & 0.67 & 0.79 & 0.92 & 0.67 & 0.73 \\ 
  SVM-L & 0.81 & 0.92 & 0.64 & 0.70 & 0.80 & 0.92 & 0.66 & 0.75 & 0.92 & 0.92 & 0.67 & 0.75 & 0.80 & 0.91 & 0.66 & 0.70 & 0.81 & 0.92 & 0.68 & 0.66 & 0.80 & 0.92 & 0.66 & 0.73 \\ 
  SVM-R & 0.79 & 0.92 & 0.68 & 0.50 & 0.81 & 0.93 & 0.70 & 0.77 & 0.93 & 0.93 & 0.67 & 0.72 & 0.70 & 0.90 & 0.66 & 0.50 & 0.74 & 0.93 & 0.70 & 0.68 & 0.71 & 0.92 & 0.65 & 0.68 \\ 
  KNN & 0.71 & 0.92 & 0.70 & 0.59 & 0.71 & 0.92 & 0.71 & 0.66 & 0.92 & 0.92 & 0.56 & 0.52 & 0.58 & 0.80 & 0.58 & 0.59 & 0.60 & 0.93 & 0.63 & 0.61 & 0.57 & 0.92 & 0.61 & 0.58 \\ 
  ANN & 0.80 & 0.92 & 0.73 & 0.77 & 0.80 & 0.89 & 0.68 & 0.78 & 0.83 & 0.83 & 0.72 & 0.75 & 0.71 & 0.88 & 0.64 & 0.77 & 0.80 & 0.93 & 0.71 & 0.72 & 0.71 & 0.92 & 0.67 & 0.77 \\ 
  CART & 0.74 & 0.86 & 0.64 & 0.60 & 0.72 & 0.86 & 0.71 & 0.78 & 0.86 & 0.86 & 0.69 & 0.78 & 0.72 & 0.90 & 0.68 & 0.60 & 0.70 & 0.86 & 0.67 & 0.50 & 0.74 & 0.89 & 0.72 & 0.71 \\ 
  NB & 0.78 & 0.86 & 0.70 & 0.67 & 0.79 & 0.85 & 0.70 & 0.71 & 0.86 & 0.86 & 0.70 & 0.65 & 0.78 & 0.85 & 0.69 & 0.67 & 0.79 & 0.91 & 0.72 & 0.74 & 0.79 & 0.86 & 0.67 & 0.68 \\ 
  \hline
  Bag-CT & 0.78 & 0.94 & 0.71 & 0.77 & 0.77 & 0.93 & 0.72 & 0.79 & 0.94 & 0.94 & 0.70 & 0.74 & 0.78 & 0.92 & 0.69 & 0.77 & 0.77 & 0.93 & 0.69 & 0.53 & 0.76 & 0.92 & 0.71 & 0.77 \\ 
  Boost-CT & 0.82 & 0.94 & 0.76 & 0.80 & 0.81 & 0.93 & 0.76 & 0.80 & 0.93 & 0.93 & 0.68 & 0.73 & 0.82 & 0.92 & 0.71 & 0.80 & 0.69 & 0.91 & 0.69 & 0.66 & 0.80 & 0.93 & 0.69 & 0.77 \\ 
  AdaBoost & 0.83 & 0.94 & 0.76 & 0.80 & 0.81 & 0.94 & 0.76 & 0.81 & 0.93 & 0.93 & 0.73 & 0.76 & 0.76 & 0.93 & 0.70 & 0.80 & 0.74 & 0.91 & 0.71 & 0.65 & 0.80 & 0.92 & 0.67 & 0.77 \\ 
  SGB & 0.83 & 0.93 & 0.76 & 0.79 & 0.82 & 0.93 & 0.75 & 0.81 & 0.94 & 0.94 & 0.72 & 0.78 & 0.82 & 0.93 & 0.70 & 0.79 & 0.80 & 0.92 & 0.71 & 0.75 & 0.79 & 0.93 & 0.70 & 0.76 \\ 
  RF & 0.81 & 0.94 & 0.73 & 0.80 & 0.81 & 0.94 & 0.75 & 0.80 & 0.94 & 0.94 & 0.73 & 0.79 & 0.81 & 0.93 & 0.73 & 0.80 & 0.80 & 0.91 & 0.72 & 0.60 & 0.81 & 0.93 & 0.73 & 0.80 \\ 
  parRF & 0.82 & 0.94 & 0.72 & 0.79 & 0.81 & 0.93 & 0.75 & 0.81 & 0.94 & 0.94 & 0.74 & 0.79 & 0.81 & 0.93 & 0.73 & 0.79 & 0.79 & 0.92 & 0.72 & 0.59 & 0.81 & 0.94 & 0.73 & 0.80 \\ 
  rotRF & 0.77 & 0.92 & 0.65 & 0.50 & 0.80 & 0.86 & 0.67 & 0.79 & 0.94 & 0.94 & 0.74 & 0.75 & 0.78 & 0.94 & 0.72 & 0.50 & 0.74 & 0.91 & 0.72 & 0.60 & 0.75 & 0.92 & 0.72 & 0.78 \\ 
  avNNet & 0.78 & 0.91 & 0.72 & 0.77 & 0.79 & 0.90 & 0.73 & 0.77 & 0.92 & 0.92 & 0.72 & 0.76 & 0.80 & 0.93 & 0.70 & 0.77 & 0.80 & 0.93 & 0.71 & 0.75 & 0.76 & 0.93 & 0.71 & 0.75 \\ 
  Logit-Boost & 0.74 & 0.90 & 0.69 & 0.68 & 0.66 & 0.93 & 0.67 & 0.71 & 0.91 & 0.91 & 0.70 & 0.76 & 0.71 & 0.91 & 0.66 & 0.68 & 0.71 & 0.90 & 0.69 & 0.63 & 0.74 & 0.90 & 0.64 & 0.78 \\ 
  bagFDA & 0.82 & 0.94 & 0.75 & 0.76 & 0.82 & 0.93 & 0.75 & 0.79 & 0.94 & 0.94 & 0.74 & 0.78 & 0.80 & 0.93 & 0.71 & 0.76 & 0.80 & 0.92 & 0.71 & 0.56 & 0.80 & 0.94 & 0.71 & 0.78 \\ 
  \hline
  AvgS & 0.82 & 0.94 & 0.75 & 0.80 & 0.81 & 0.94 & 0.75 & 0.81 & 0.94 & 0.94 & 0.74 & 0.80 & 0.81 & 0.94 & 0.73 & 0.80 & 0.79 & 0.93 & 0.73 & 0.69 & 0.80 & 0.94 & 0.73 & 0.78 \\ 
  AvgW & 0.82 & 0.94 & 0.75 & 0.80 & 0.81 & 0.94 & 0.75 & 0.81 & 0.94 & 0.94 & 0.74 & 0.80 & 0.81 & 0.94 & 0.73 & 0.80 & 0.79 & 0.92 & 0.73 & 0.69 & 0.80 & 0.94 & 0.73 & 0.78 \\ 
  Stack & 0.81 & 0.94 & 0.73 & 0.80 & 0.78 & 0.95 & 0.73 & 0.79 & 0.93 & 0.93 & 0.71 & 0.52 & 0.73 & 0.92 & 0.68 & 0.80 & 0.76 & 0.93 & 0.71 & 0.71 & 0.77 & 0.95 & 0.68 & 0.80 \\ 
  \\[-1.8ex]\hline 
\hline \\[-1.8ex]
\end{tabular}}
\end{sidewaystable}

\begin{sidewaystable}
\caption{Results for HM per model and data set}
\label{results:details_HM}
\scalebox{0.75}{
\begin{tabular}{lrrrr|rrrr|rrrr|rrrr|rrrr|rrrr}
 \\[-1.8ex]\hline 
\hline \\[-1.8ex]
  Data set & \multicolumn{4}{c|}{original} & \multicolumn{4}{c|}{down-sampling} & \multicolumn{4}{c|}{up-sampling} & \multicolumn{4}{c|}{SMOTE} & \multicolumn{4}{c|}{ROSE} & \multicolumn{4}{c}{BSMOTE}\\
 & GC & AC & TC & GMSC & GC & AC & TC & GMSC & GC & AC & TC & GMSC & GC & AC & TC & GMSC & GC & AC & TC & GMSC & GC & AC & TC & GMSC \\ 
  \hline
LDA & 0.34 & 0.68 & 0.16 & 0.06 & 0.32 & 0.66 & 0.14 & 0.04 & 0.68 & 0.68 & 0.16 & 0.04 & 0.35 & 0.67 & 0.16 & 0.06 & 0.33 & 0.71 & 0.19 & 0.04 & 0.32 & 0.67 & 0.14 & 0.03 \\ 
  QDA & 0.27 & 0.63 & 0.13 & 0.08 & 0.27 & 0.62 & 0.12 & 0.13 & 0.62 & 0.62 & 0.12 & 0.08 & 0.26 & 0.61 & 0.13 & 0.08 & 0.27 & 0.65 & 0.15 & 0.09 & 0.24 & 0.62 & 0.13 & 0.08 \\ 
  FDA & 0.28 & 0.74 & 0.15 & 0.12 & 0.19 & 0.68 & 0.19 & 0.14 & 0.71 & 0.71 & 0.16 & 0.12 & 0.25 & 0.71 & 0.17 & 0.12 & 0.26 & 0.63 & 0.18 & 0.01 & 0.26 & 0.64 & 0.17 & 0.13 \\ 
  LogReg & 0.34 & 0.64 & 0.16 & 0.04 & 0.32 & 0.66 & 0.14 & 0.13 & 0.66 & 0.66 & 0.16 & 0.12 & 0.35 & 0.63 & 0.16 & 0.04 & 0.32 & 0.72 & 0.19 & 0.04 & 0.31 & 0.65 & 0.13 & 0.11 \\ 
  SVM-L & 0.36 & 0.68 & 0.16 & 0.11 & 0.32 & 0.67 & 0.13 & 0.13 & 0.69 & 0.69 & 0.14 & 0.13 & 0.34 & 0.65 & 0.15 & 0.11 & 0.35 & 0.73 & 0.19 & 0.04 & 0.32 & 0.66 & 0.12 & 0.12 \\ 
  SVM-R & 0.30 & 0.71 & 0.17 & 0.01 & 0.33 & 0.72 & 0.15 & 0.15 & 0.72 & 0.72 & 0.10 & 0.12 & 0.17 & 0.57 & 0.11 & 0.01 & 0.22 & 0.71 & 0.17 & 0.08 & 0.20 & 0.64 & 0.08 & 0.11 \\ 
  KNN & 0.14 & 0.65 & 0.14 & 0.03 & 0.16 & 0.65 & 0.14 & 0.03 & 0.67 & 0.67 & 0.02 & 0.00 & 0.03 & 0.41 & 0.02 & 0.03 & 0.04 & 0.66 & 0.05 & 0.03 & 0.02 & 0.70 & 0.04 & 0.04 \\ 
  ANN & 0.36 & 0.67 & 0.19 & 0.14 & 0.33 & 0.62 & 0.12 & 0.15 & 0.49 & 0.49 & 0.17 & 0.13 & 0.21 & 0.60 & 0.10 & 0.14 & 0.33 & 0.72 & 0.18 & 0.12 & 0.17 & 0.66 & 0.11 & 0.14 \\ 
  CART & 0.22 & 0.57 & 0.12 & 0.07 & 0.20 & 0.57 & 0.15 & 0.12 & 0.57 & 0.57 & 0.16 & 0.12 & 0.18 & 0.65 & 0.14 & 0.07 & 0.15 & 0.57 & 0.14 & 0.00 & 0.20 & 0.59 & 0.15 & 0.09 \\ 
  NB & 0.29 & 0.55 & 0.16 & 0.09 & 0.29 & 0.54 & 0.15 & 0.07 & 0.56 & 0.56 & 0.16 & 0.08 & 0.27 & 0.52 & 0.15 & 0.09 & 0.29 & 0.64 & 0.19 & 0.11 & 0.29 & 0.56 & 0.11 & 0.07 \\ 
  \hline
  Bag-CT & 0.30 & 0.71 & 0.12 & 0.13 & 0.24 & 0.67 & 0.16 & 0.15 & 0.71 & 0.71 & 0.14 & 0.12 & 0.28 & 0.60 & 0.12 & 0.13 & 0.23 & 0.66 & 0.14 & 0.00 & 0.24 & 0.65 & 0.13 & 0.13 \\ 
  Boost-CT & 0.37 & 0.72 & 0.20 & 0.17 & 0.34 & 0.71 & 0.22 & 0.17 & 0.72 & 0.72 & 0.13 & 0.14 & 0.34 & 0.70 & 0.16 & 0.17 & 0.16 & 0.63 & 0.16 & 0.05 & 0.35 & 0.69 & 0.13 & 0.16 \\ 
  AdaBoost & 0.39 & 0.71 & 0.21 & 0.17 & 0.34 & 0.71 & 0.20 & 0.16 & 0.69 & 0.69 & 0.18 & 0.14 & 0.27 & 0.64 & 0.14 & 0.17 & 0.23 & 0.61 & 0.19 & 0.06 & 0.32 & 0.63 & 0.11 & 0.16 \\ 
  SGB & 0.41 & 0.70 & 0.22 & 0.16 & 0.37 & 0.69 & 0.21 & 0.16 & 0.71 & 0.71 & 0.17 & 0.15 & 0.37 & 0.67 & 0.15 & 0.16 & 0.32 & 0.66 & 0.19 & 0.10 & 0.29 & 0.69 & 0.13 & 0.14 \\ 
  RF & 0.36 & 0.72 & 0.16 & 0.17 & 0.33 & 0.71 & 0.20 & 0.17 & 0.72 & 0.72 & 0.18 & 0.15 & 0.33 & 0.67 & 0.18 & 0.17 & 0.31 & 0.66 & 0.19 & 0.02 & 0.35 & 0.71 & 0.17 & 0.17 \\ 
  parRF & 0.36 & 0.73 & 0.16 & 0.17 & 0.35 & 0.71 & 0.21 & 0.18 & 0.71 & 0.71 & 0.18 & 0.16 & 0.34 & 0.65 & 0.17 & 0.17 & 0.29 & 0.68 & 0.17 & 0.03 & 0.35 & 0.69 & 0.17 & 0.17 \\ 
  rotRF & 0.26 & 0.69 & 0.14 & 0.00 & 0.32 & 0.57 & 0.14 & 0.14 & 0.71 & 0.71 & 0.19 & 0.15 & 0.29 & 0.71 & 0.15 & 0.00 & 0.22 & 0.64 & 0.19 & 0.05 & 0.23 & 0.66 & 0.15 & 0.14 \\ 
  avNNet & 0.30 & 0.65 & 0.18 & 0.14 & 0.32 & 0.64 & 0.19 & 0.15 & 0.65 & 0.65 & 0.16 & 0.14 & 0.30 & 0.66 & 0.15 & 0.14 & 0.32 & 0.72 & 0.18 & 0.12 & 0.26 & 0.66 & 0.16 & 0.12 \\ 
  Logit-Boost & 0.19 & 0.59 & 0.14 & 0.08 & 0.10 & 0.66 & 0.09 & 0.08 & 0.61 & 0.61 & 0.14 & 0.11 & 0.16 & 0.60 & 0.10 & 0.08 & 0.14 & 0.61 & 0.16 & 0.05 & 0.19 & 0.58 & 0.07 & 0.12 \\ 
  bagFDA & 0.36 & 0.69 & 0.20 & 0.14 & 0.36 & 0.68 & 0.20 & 0.16 & 0.70 & 0.70 & 0.21 & 0.15 & 0.34 & 0.67 & 0.18 & 0.14 & 0.32 & 0.67 & 0.20 & 0.01 & 0.33 & 0.71 & 0.18 & 0.15 \\ 
  \hline
  AvgS & 0.39 & 0.71 & 0.21 & 0.19 & 0.37 & 0.71 & 0.20 & 0.17 & 0.70 & 0.70 & 0.19 & 0.17 & 0.35 & 0.71 & 0.19 & 0.19 & 0.31 & 0.71 & 0.21 & 0.09 & 0.34 & 0.71 & 0.18 & 0.16 \\ 
  AvgW & 0.39 & 0.71 & 0.21 & 0.19 & 0.37 & 0.71 & 0.20 & 0.17 & 0.71 & 0.71 & 0.19 & 0.17 & 0.34 & 0.71 & 0.19 & 0.19 & 0.31 & 0.71 & 0.20 & 0.09 & 0.34 & 0.71 & 0.18 & 0.16 \\ 
  Stack & 0.37 & 0.71 & 0.20 & 0.17 & 0.30 & 0.72 & 0.20 & 0.14 & 0.70 & 0.70 & 0.17 & 0.06 & 0.22 & 0.65 & 0.13 & 0.17 & 0.25 & 0.66 & 0.17 & 0.07 & 0.30 & 0.71 & 0.11 & 0.16 \\ 
   \\[-1.8ex]\hline 
\hline \\[-1.8ex]
\end{tabular}}
\end{sidewaystable}

\begin{sidewaystable}
\caption{Results for KS per model and data set}
\label{results:details_KS}
\scalebox{0.75}{
\begin{tabular}{lrrrr|rrrr|rrrr|rrrr|rrrr|rrrr}
 \\[-1.8ex]\hline 
\hline \\[-1.8ex]
  Data set & \multicolumn{4}{c|}{original} & \multicolumn{4}{c|}{down-sampling} & \multicolumn{4}{c|}{up-sampling} & \multicolumn{4}{c|}{SMOTE} & \multicolumn{4}{c|}{ROSE} & \multicolumn{4}{c}{BSMOTE}\\
 & GC & AC & TC & GMSC & GC & AC & TC & GMSC & GC & AC & TC & GMSC & GC & AC & TC & GMSC & GC & AC & TC & GMSC & GC & AC & TC & GMSC \\ 
  \hline
LDA & 0.50 & 0.76 & 0.34 & 0.29 & 0.48 & 0.74 & 0.30 & 0.26 & 0.76 & 0.76 & 0.34 & 0.28 & 0.50 & 0.75 & 0.33 & 0.29 & 0.47 & 0.80 & 0.36 & 0.27 & 0.49 & 0.76 & 0.32 & 0.24 \\ 
  QDA & 0.41 & 0.75 & 0.33 & 0.33 & 0.40 & 0.73 & 0.33 & 0.43 & 0.73 & 0.73 & 0.32 & 0.32 & 0.41 & 0.71 & 0.31 & 0.33 & 0.38 & 0.77 & 0.33 & 0.35 & 0.39 & 0.73 & 0.32 & 0.29 \\ 
  FDA & 0.45 & 0.79 & 0.31 & 0.39 & 0.38 & 0.77 & 0.37 & 0.44 & 0.78 & 0.78 & 0.32 & 0.41 & 0.42 & 0.78 & 0.34 & 0.39 & 0.40 & 0.74 & 0.36 & 0.11 & 0.40 & 0.72 & 0.36 & 0.42 \\ 
  LogReg & 0.48 & 0.74 & 0.32 & 0.27 & 0.49 & 0.76 & 0.30 & 0.38 & 0.74 & 0.74 & 0.32 & 0.38 & 0.51 & 0.73 & 0.30 & 0.27 & 0.47 & 0.81 & 0.36 & 0.28 & 0.45 & 0.75 & 0.30 & 0.36 \\ 
  SVM-L & 0.51 & 0.77 & 0.30 & 0.34 & 0.48 & 0.75 & 0.29 & 0.37 & 0.78 & 0.78 & 0.32 & 0.37 & 0.52 & 0.75 & 0.31 & 0.34 & 0.51 & 0.82 & 0.35 & 0.29 & 0.46 & 0.76 & 0.29 & 0.38 \\ 
  SVM-R & 0.46 & 0.80 & 0.33 & 0.10 & 0.46 & 0.81 & 0.33 & 0.42 & 0.81 & 0.81 & 0.30 & 0.37 & 0.35 & 0.67 & 0.23 & 0.10 & 0.41 & 0.80 & 0.32 & 0.30 & 0.34 & 0.75 & 0.25 & 0.33 \\ 
  KNN & 0.34 & 0.74 & 0.31 & 0.14 & 0.36 & 0.74 & 0.33 & 0.21 & 0.77 & 0.77 & 0.12 & 0.04 & 0.17 & 0.59 & 0.16 & 0.14 & 0.15 & 0.76 & 0.20 & 0.15 & 0.13 & 0.80 & 0.20 & 0.15 \\ 
  ANN & 0.51 & 0.75 & 0.37 & 0.43 & 0.50 & 0.74 & 0.31 & 0.44 & 0.64 & 0.64 & 0.34 & 0.39 & 0.39 & 0.70 & 0.26 & 0.43 & 0.49 & 0.81 & 0.36 & 0.33 & 0.37 & 0.75 & 0.29 & 0.43 \\ 
  CART & 0.41 & 0.72 & 0.28 & 0.19 & 0.45 & 0.72 & 0.33 & 0.46 & 0.72 & 0.72 & 0.37 & 0.44 & 0.41 & 0.76 & 0.33 & 0.19 & 0.37 & 0.72 & 0.33 & 0.01 & 0.41 & 0.72 & 0.35 & 0.39 \\ 
  NB & 0.44 & 0.68 & 0.35 & 0.35 & 0.43 & 0.67 & 0.34 & 0.38 & 0.68 & 0.68 & 0.36 & 0.30 & 0.45 & 0.66 & 0.28 & 0.35 & 0.48 & 0.74 & 0.37 & 0.39 & 0.47 & 0.69 & 0.26 & 0.31 \\ 
  \hline
  Bag-CT & 0.44 & 0.79 & 0.32 & 0.46 & 0.40 & 0.76 & 0.37 & 0.43 & 0.80 & 0.80 & 0.32 & 0.41 & 0.45 & 0.70 & 0.29 & 0.46 & 0.39 & 0.76 & 0.32 & 0.07 & 0.43 & 0.76 & 0.31 & 0.42 \\ 
  Boost-CT & 0.53 & 0.77 & 0.39 & 0.49 & 0.47 & 0.80 & 0.40 & 0.47 & 0.77 & 0.77 & 0.32 & 0.39 & 0.48 & 0.78 & 0.36 & 0.49 & 0.29 & 0.72 & 0.33 & 0.27 & 0.48 & 0.77 & 0.29 & 0.44 \\ 
  AdaBoost & 0.55 & 0.80 & 0.39 & 0.47 & 0.50 & 0.79 & 0.38 & 0.48 & 0.76 & 0.76 & 0.38 & 0.43 & 0.46 & 0.73 & 0.35 & 0.47 & 0.38 & 0.73 & 0.35 & 0.25 & 0.48 & 0.69 & 0.29 & 0.41 \\ 
  SGB & 0.54 & 0.77 & 0.39 & 0.48 & 0.51 & 0.78 & 0.39 & 0.50 & 0.78 & 0.78 & 0.37 & 0.46 & 0.48 & 0.76 & 0.31 & 0.48 & 0.47 & 0.76 & 0.34 & 0.40 & 0.44 & 0.76 & 0.31 & 0.43 \\ 
  RF & 0.51 & 0.77 & 0.34 & 0.48 & 0.49 & 0.77 & 0.40 & 0.50 & 0.78 & 0.78 & 0.36 & 0.45 & 0.49 & 0.75 & 0.34 & 0.48 & 0.47 & 0.73 & 0.35 & 0.19 & 0.50 & 0.80 & 0.35 & 0.47 \\ 
  parRF & 0.51 & 0.79 & 0.33 & 0.47 & 0.50 & 0.77 & 0.40 & 0.48 & 0.78 & 0.78 & 0.38 & 0.45 & 0.52 & 0.72 & 0.33 & 0.47 & 0.44 & 0.76 & 0.35 & 0.17 & 0.48 & 0.77 & 0.35 & 0.49 \\ 
  rotRF & 0.44 & 0.77 & 0.30 & 0.00 & 0.49 & 0.72 & 0.33 & 0.47 & 0.79 & 0.79 & 0.38 & 0.44 & 0.45 & 0.77 & 0.32 & 0.00 & 0.42 & 0.73 & 0.35 & 0.21 & 0.37 & 0.75 & 0.36 & 0.45 \\ 
  avNNet & 0.49 & 0.74 & 0.35 & 0.41 & 0.50 & 0.75 & 0.35 & 0.45 & 0.74 & 0.74 & 0.32 & 0.44 & 0.46 & 0.74 & 0.31 & 0.41 & 0.48 & 0.81 & 0.35 & 0.41 & 0.43 & 0.75 & 0.35 & 0.38 \\ 
  Logit-Boost & 0.38 & 0.72 & 0.30 & 0.35 & 0.24 & 0.76 & 0.28 & 0.40 & 0.73 & 0.73 & 0.33 & 0.39 & 0.34 & 0.70 & 0.24 & 0.35 & 0.30 & 0.72 & 0.34 & 0.24 & 0.37 & 0.67 & 0.22 & 0.44 \\ 
  bagFDA & 0.54 & 0.76 & 0.39 & 0.41 & 0.51 & 0.75 & 0.39 & 0.45 & 0.77 & 0.77 & 0.38 & 0.43 & 0.54 & 0.76 & 0.34 & 0.41 & 0.48 & 0.77 & 0.37 & 0.17 & 0.50 & 0.79 & 0.35 & 0.44 \\ 
  \hline
  AvgS & 0.53 & 0.78 & 0.40 & 0.49 & 0.54 & 0.79 & 0.40 & 0.48 & 0.78 & 0.78 & 0.39 & 0.46 & 0.50 & 0.78 & 0.35 & 0.49 & 0.47 & 0.80 & 0.36 & 0.30 & 0.47 & 0.79 & 0.37 & 0.46 \\ 
  AvgW & 0.53 & 0.78 & 0.41 & 0.49 & 0.54 & 0.79 & 0.40 & 0.48 & 0.78 & 0.78 & 0.39 & 0.46 & 0.50 & 0.78 & 0.35 & 0.49 & 0.47 & 0.79 & 0.36 & 0.30 & 0.47 & 0.79 & 0.36 & 0.46 \\ 
  Stack & 0.50 & 0.79 & 0.36 & 0.48 & 0.51 & 0.79 & 0.40 & 0.49 & 0.79 & 0.79 & 0.41 & 0.16 & 0.38 & 0.73 & 0.29 & 0.48 & 0.43 & 0.76 & 0.36 & 0.37 & 0.50 & 0.77 & 0.32 & 0.47 \\
   \\[-1.8ex]\hline 
\hline \\[-1.8ex]
\end{tabular}}
\end{sidewaystable}

\end{appendices}

\end{document}